\documentclass[twocolumn,showpacs,showkeys,preprintnumbers,amsmath,amssymb,floatfix,aps,prc,10pt]{revtex4-1}
\usepackage[latin9]{inputenc}

\usepackage{feynmp-auto}
\usepackage{color}
\usepackage{graphicx}
\usepackage{mathtools}
\usepackage{makecell} 
\usepackage{multirow}
\usepackage{siunitx}
\usepackage{array,booktabs,tabularx,longtable}
\usepackage{afterpage} 
\usepackage[running]{lineno}
\usepackage[colorlinks,urlcolor=blue]{hyperref}
\usepackage{breakurl}

\usepackage[normalem]{ulem}
\newcounter{univ_counter}
\addtocounter{univ_counter} {1}
\edef\JLAB{$^{\arabic{univ_counter}}$ }

\addtocounter{univ_counter} {1} 
\edef\SCAROLINA{$^{\arabic{univ_counter}}$ }

\addtocounter{univ_counter} {1} 
\edef\MSU{$^{\arabic{univ_counter}}$ }

\addtocounter{univ_counter} {1} 
\edef\UCONN{$^{\arabic{univ_counter}}$ }

\begin{document}

\preprint{Phys. Rev. C}

\title{First Results on Nucleon Resonance Electroexcitation Amplitudes from $ep \to e'\pi^+\pi^-p'$ Cross Sections at $W$ from $1.56-1.76$~GeV and $Q^2$ from $2.0-5.0$~GeV$^2$}

\newcommand{\orcidauthorA}{0000-0002-1280-0983} 
\newcommand{\orcidauthorB}{0000-0001-9833-3695} 
\newcommand{\orcidauthorC}{0000-0002-4557-1320} 

\author{
V.I.~Mokeev,\JLAB\,\!\!\SCAROLINA\
P.~Achenbach,\JLAB\
V.D.~Burkert,\JLAB\
D.S.~Carman,\JLAB\
R.W.~Gothe,\SCAROLINA\
E.L.~Isupov,\MSU\
K.~Joo,\UCONN\
K.~Neupane,\SCAROLINA\
Y.~Wunderlich\UCONN\
}

\affiliation{\JLAB Thomas Jefferson National Accelerator Facility, Newport News, Virginia 23606}
\affiliation{\SCAROLINA University of South Carolina, Columbia, South Carolina 29208}
\affiliation{\MSU Skobeltsyn Institute of Nuclear Physics and Physics Department at Lomonosov Moscow State University, 119234 Moscow, Russia}
\affiliation{\UCONN University of Connecticut, Storrs, Connecticut 06269}

\date{\today}

\begin{abstract}
The first results on the electroexcitation amplitudes or the $\gamma_vpN^*$ electrocouplings for nucleon resonances ($N^*$s) in the third resonance region are presented. They were obtained from $\pi^+\pi^-p$ electroproduction differential cross sections measured with the CLAS detector and analyzed using the Jefferson Lab-Moscow State University (JM) reaction model. The analysis covers the invariant mass range of the final-state hadrons $W$ from 1.56 to 1.76~GeV and virtual photon four-momentum squared $Q^2$ from 2.0 to 5.0~GeV$^2$. Consistent results on the electroexcitation amplitudes of the $N(1675)5/2^-$ and $N(1680)5/2^+$ obtained from independent analyses of both $\pi N$ and $\pi^+\pi^-p$ final states, demonstrate the capability of reaction models to extract the $\gamma_v p N^*$ electrocouplings for $N^*$s in the third resonance region. Also, for the first time, the electrocouplings of the $\Delta(1700)3/2^-$ and $N(1720)3/2^+$, which predominantly decay into $\pi\pi N$ final states, have become available for $Q^2 > 2.0$~GeV$^2$. Finally, contributions from a new $N'(1720)3/2^+$ baryon state to the $\pi^+\pi^-p$ differential cross sections have been observed for $Q^2 < 5.0$~GeV$^2$. The new results on resonance electrocouplings in the third resonance region offer new opportunities to explore various aspects of the strong QCD regime responsible for the generation of nucleon excited states, in particular, shedding light on the emergence of hadron mass in connection with dynamical chiral symmetry breaking.  
\end{abstract}

\maketitle
\noindent
PACS: 13.40.-f, 14.20.Gk, 12.40.Nn \\
Keywords: CLAS, electron scattering, resonance contributions, parton distributions

\section{Introduction}
\label{sec:intro}

Extensive studies of exclusive meson electroproduction with the CLAS detector at Jefferson Lab (JLab) during the 6-GeV era have provided the dominant share of the world's available information on most exclusive meson electroproduction channels in the nucleon resonance region \cite{Burkert:2019kxy, Mokeev:2022xfo, Brodsky:2020vco}. For the first time, a large body of data on differential cross sections and polarization asymmetries have become available within the nearly full acceptance coverage of the CLAS detector~\cite{CLAS:2003umf}.

Nucleon resonance ($N^*$) electroexcitation amplitudes, also known as $\gamma_v p N^*$ electrocouplings, have been extracted from analyses of CLAS data for most excited nucleon states with masses up to 1.8~GeV and for virtual photon four-momentum squared $Q^2$ up to 5.0~GeV$^2$~\cite{Mokeev:2022xfo, Burkert:2019kxy, Carman:2020qmb}. These electrocouplings have been independently determined from studies of $\pi^0 p$ and $\pi^+ n$ (collectively referred to as $\pi N$) exclusive channels \cite{Aznauryan:2011qj, Aznauryan:2002gd, Aznauryan:2009mx, Park:2014yea, Tiator:2011pw} for $Q^2 < 5.0$~GeV$^2$ for the resonances across the mass range up to 1.8~GeV and in $\eta p$ \cite{CLAS:2007bvs} across the mass range up to 1.6~GeV for $Q^2 < 3.0$~GeV$^2$. Electrocouplings of the resonances located in the mass range below 1.6~GeV have been determined also from $\pi^+ \pi^- p$ electroproduction for $Q^2 < 5.0$~GeV$^2$, while in the mass range from 1.6~GeV to 1.8~GeV they were obtained within a more limited coverage over $Q^2 < 1.5$~GeV$^2$ only~\cite{Mokeev:2008iw, Mokeev:2012vsa, Mokeev:2015lda, Mokeev:2023zhq}.

Recently, first results on the $\gamma_v p N^*$ electrocouplings were obtained from a global coupled-channel analysis of $\pi N$, $\eta p$, $K\Lambda$, and $K\Sigma$ final states in photo-, electro-, and hadroproduction reactions. This effort, developed jointly by the J\"ulich-Bonn-George Washington groups, yielded electrocouplings for resonances in the mass range up to 1.8~GeV for $Q^2 < 5.0$~GeV$^2$~\cite{Wang:2024byt}. Earlier, the Argonne-Osaka coupled-channel analysis provided results on the electrocouplings of the $\Delta(1232)3/2^+$ and $N(1440)1/2^+$ \cite{Kamano:2018sfb}.

The electrocoupling results provide unique information that enables exploration of many facets of the strong interaction in the regime where the QCD running coupling $\alpha_s$ is large--comparable to unity, known as the strongly coupled or sQCD regime~\cite{Cui:2019dwv}. This can be achieved through studies of prominent $N^*$ states with diverse quantum numbers and structural features. These results also allow us to investigate the evolution of nucleon resonance structure with distance (or $Q^2$) across the transition from the sQCD to the perturbative QCD (pQCD) regimes. Furthermore, knowledge of the electrocouplings over a broad range of $W$ and $Q^2$ makes it possible to isolate the resonant contributions to inclusive structure functions, offering a new opportunity to shed light on the evolution of partonic degrees of freedom in the structure of the ground-state nucleon at large parton fractional momenta $x$ in the resonance region spanning the transition from the sQCD to pQCD regimes. This work is currently in progress \cite{HillerBlin:2019jgp,Blin:2021twt,HillerBlin:2022ltm} and requires extending our knowledge of $\gamma_v p N^*$ electrocouplings, particularly in the region $1.6 < W < 2.0$~GeV.

Analyses of the CLAS/JLab results on the $Q^2$-evolution of $\gamma_v p N^*$ electrocouplings, carried out using Continuum Schwinger Methods (CSMs) under a traceable connection to the QCD Lagrangian \cite{Burkert:2017djo,Segovia:2014aza,Segovia:2015hra}, and upported by the experimental results on $\gamma_vpN^*$ electrocouplings~\cite{Aznauryan:2009mx,Park:2014yea,CLAS:2008ihz, CLAS:2002xbv} as well as by the quark models \cite{Aznauryan:2014xea, Aznauryan:2018okk,Ramalho:2023hqd, Sirca:2026bwv, Obukhovsky:2019xrs}, have demonstrated that the structure of $N^*$ states is governed by a complex interplay between an inner core of three dressed quarks and an external meson-baryon cloud. At distance scales corresponding to $Q^2 > 2$~GeV$^2$, the three-quark core becomes the biggest contributor to $N^*$ structure. Advances in CSMs have conclusively shown that, at these distances, $\gamma_v p N^*$ electrocouplings provide a novel avenue for elucidating the mechanism responsible for the generation of more than 98\% of hadron mass \cite{Ding:2022ows, Carman:2023zke, Achenbach:2025kfx}, thereby addressing one of the key open problems in the Standard Model: the emergence of hadron mass (EHM).

The $\pi N$ and $\pi^+\pi^-p$ electroproduction channels are the two biggest contributors in the resonance region. The low-lying $N^*$ states with masses below 1.6~GeV decay predominantly into $\pi N$ final states, making single-pion electroproduction data the primary source of information on their electrocouplings~\cite{Aznauryan:2011qj, Park:2014yea, Aznauryan:2009mx}. At the same time, the branching fraction (BF) for these resonances decaying into $\pi\pi N$ final states remains significant--on the order of 40\%--allowing for an independent determination of their electrocouplings from this channel~\cite{Mokeev:2012vsa, Mokeev:2015lda}. Consistent results on the electrocouplings from independent analyses of both $\pi N$ and $\pi^+\pi^-p$ channels demonstrate the reliability of the $\pi N$  and $\pi^+\pi^-p$ reaction models for extracting these quantities and for evaluating the systematic uncertainties associated with their determination~\cite{Aznauryan:2005tp, Mokeev:2015lda, Achenbach:2025kfx}.

The first results on the $\pi^+\pi^-p$ differential cross sections for $W < 2.0$~GeV and $Q^2 = 2.0 - 5.0$~GeV$^2$ were published in Ref.~\cite{CLAS:2017fja}. In the recent CLAS publication, Ref.~\cite{Trivedi:2026fsd}, the evaluation of the $\pi^+\pi^-p$ differential cross sections has been substantially improved through the development of a new approach for determining the $\pi^+\pi^-p$ event detection efficiency, accounting for contributions from regions of the reaction phase space with low detection efficiency or regions outside the detector acceptance. For the extraction of the $\gamma_v p N^*$ electrocouplings presented in this paper, we analyzed the $\pi^+\pi^-p$ differential cross sections reported in Ref.~\cite{Trivedi:2026fsd}.

Currently, studies of $\pi^+\pi^-p$ electroproduction have provided information on the $\gamma_v p N^*$ electrocouplings of the $N(1440)1/2^+$ and $N(1520)3/2^-$ for $Q^2$ from $0.2-5.0$~GeV$^2$ \cite{Brodsky:2020vco,Burkert:2019kxy} determined from $\pi^+\pi^-p$ differential cross sections for $Q^2 = 0.2 - 1.5$~GeV$^2$ in Refs.~\cite{CLAS:2008ihz, CLAS:2002xbv}, and for $Q^2 = 2.0 - 5.0$~GeV$^2$ in Ref.~\cite{Trivedi:2026fsd}. The electrocouplings of the $\Delta(1600)3/2^+$ for $Q^2$ from $2.0-5.0$~GeV$^2$ were determined \cite{Mokeev:2023zhq} from $\pi^+\pi^-p$ data in Ref.~\cite{Trivedi:2026fsd}. The results on the electrocouplings of the $\Delta(1600)3/2^+$, analyzed within the CSM, were of particular importance for validating insight into EHM from the experimental results on the $Q^2$-evolution of the electrocouplings~\cite{Ding:2022ows,Carman:2023zke,Achenbach:2025kfx}.

In the invariant mass region $W > 1.6$~GeV, the $\Delta(1600)3/2^+$, $\Delta(1700)3/2^-$, and $N(1720)3/2^+$ decay predominantly into $\pi\pi N$ final states. In addition, a new baryon state, the $N'(1720)3/2^+$, has been observed in combined studies of $\pi^+\pi^-p$ photo- and electroproduction data~\cite{Mokeev:2020hhu}. This resonance decays primarily into $\pi\pi N$ final states. It is currently the only new baryon state--previously classified as a ``missing" resonance--for which experimental information on the $Q^2$-evolution of its electrocouplings is available. These results offer new insights into the structural features of such ``missing" resonances, which may explain why their observation has remained elusive in photo- and hadroproduction analyses for such a long time.

Hence, studies of $\pi^+\pi^-p$ electroproduction at $W > 1.6$~GeV appear particularly promising for extending information on $\gamma_v p N^*$ electrocouplings of resonances in the third resonance region, in particular, for the $\Delta(1700)3/2^-$, $N(1720)3/2^+$, and $N'(1720)3/2^+$. A comparison of the $N(1675)5/2^-$ and $N(1680)5/2^+$ electrocouplings, available both from $\pi N$ and $\pi^+\pi^-p$ electroproduction, offers a nearly model-independent test of the capabilities of reaction models to extract electrocouplings for $N^*$ states in the third resonance region from independent analyses of these exclusive channels. The analysis of $\pi^+\pi^-p$ data in this mass range also provides an opportunity to explore the manifestation of the new $N'(1720)3/2^+$ at $2.0 < Q^2 < 5.0$~GeV$^2$, and to gain insight into its internal structure.

The extraction of the $\gamma_v p N^*$ electrocouplings for nucleon resonances in the third resonance region for $Q^2$ from $2.0-5.0$~GeV$^2$ from $\pi^+\pi^-p$ electroproduction data represents the focus of this work. The paper is organized as follows: Section~\ref{data_model} provides an overview of the CLAS $\pi^+\pi^-p$ differential cross sections and reviews the capability of the JM23 model~\cite{Mokeev:2023zhq} relied on for the extraction of the resonance parameters to describe the data. Section~\ref{fit_strategy} details the procedures for extraction of the resonance parameters from the cross section fits. The resonance electrocouplings, masses, and total and partial decay widths to the $\pi\Delta$ and $\rho p$ final states for the $\Delta(1600)3/2^+$, $N(1675)5/2^-$, $N(1680)5/2^+$, $\Delta(1700)3/2^-$, $N(1720)3/2^+$, and $N'(1720)3/2^+$, along with their impact on the exploration of the strong interaction dynamics that is responsible for generating $N^*$ structure is presented and discussed in Section~\ref{elcoupl_hadrdec}. Finally, Section~\ref{concl_outlook} presents our conclusions from this work and an outlook for the future.

\section{Electroproduction Cross Sections and Reaction Model}
\label{data_model}

The $\gamma_vpN^*$ electrocouplings of the excited nucleon states $\Delta(1600)3/2^+$, $N(1675)5/2^-$, $N(1680)5/2^+$, $\Delta(1700)3/2^-$, $N(1720)3/2^+$, and $N'(1720)3/2^+$ were deduced from fits of the $\pi^+\pi^-p$ differential cross sections within the overlapping $(W,Q^2)$-regions listed in Table~\ref{wq2bins}. Each region was fit independently. In this section we describe the $\pi^+\pi^-p$ differential cross sections measured with CLAS for $W$ from $1.56-1.76$~GeV and $Q^2$ from $2.0-5.0$~GeV$^2$~\cite{Trivedi:2026fsd} that were used for the extraction of the resonance parameters. We also present the basic features of the JM model relevant for the extraction of the electrocouplings from the $\pi^+\pi^-p$ data in the third resonance region.

\begin{table}
\begin{center}
\begin{tabular}{|c|c|c|} \hline
                   & $Q^2$-Interval    & $Q^2$-Interval    \\
                   & $2.0-3.5$ GeV$^2$ & $3.0-5.0$ GeV$^2$ \\ \hline
                   & $1.56-1.66$       & $1.56-1.66$       \\
$W$-interval (GeV) & $1.61-1.71$       & $1.61-1.71$       \\
                   & $1.66-1.76$       & $1.66-1.71$       \\ \hline
\end{tabular}
\caption{Kinematic regions of the CLAS $\pi^+\pi^-p$ differential cross section data \cite{Trivedi:2026fsd} analyzed for extraction of the $\Delta(1600)3/2^+$, $N(1675)5/2^-$, $N(1680)5/2^+$, $\Delta(1700)3/2^-$, $N(1720)3/2^+$, and $N'(1720)3/2^+$ electrocouplings.}
\label{wq2bins}
\end{center}
\end{table}

\begin{figure}[htbp]
\begin{center}
\includegraphics[width=8cm]{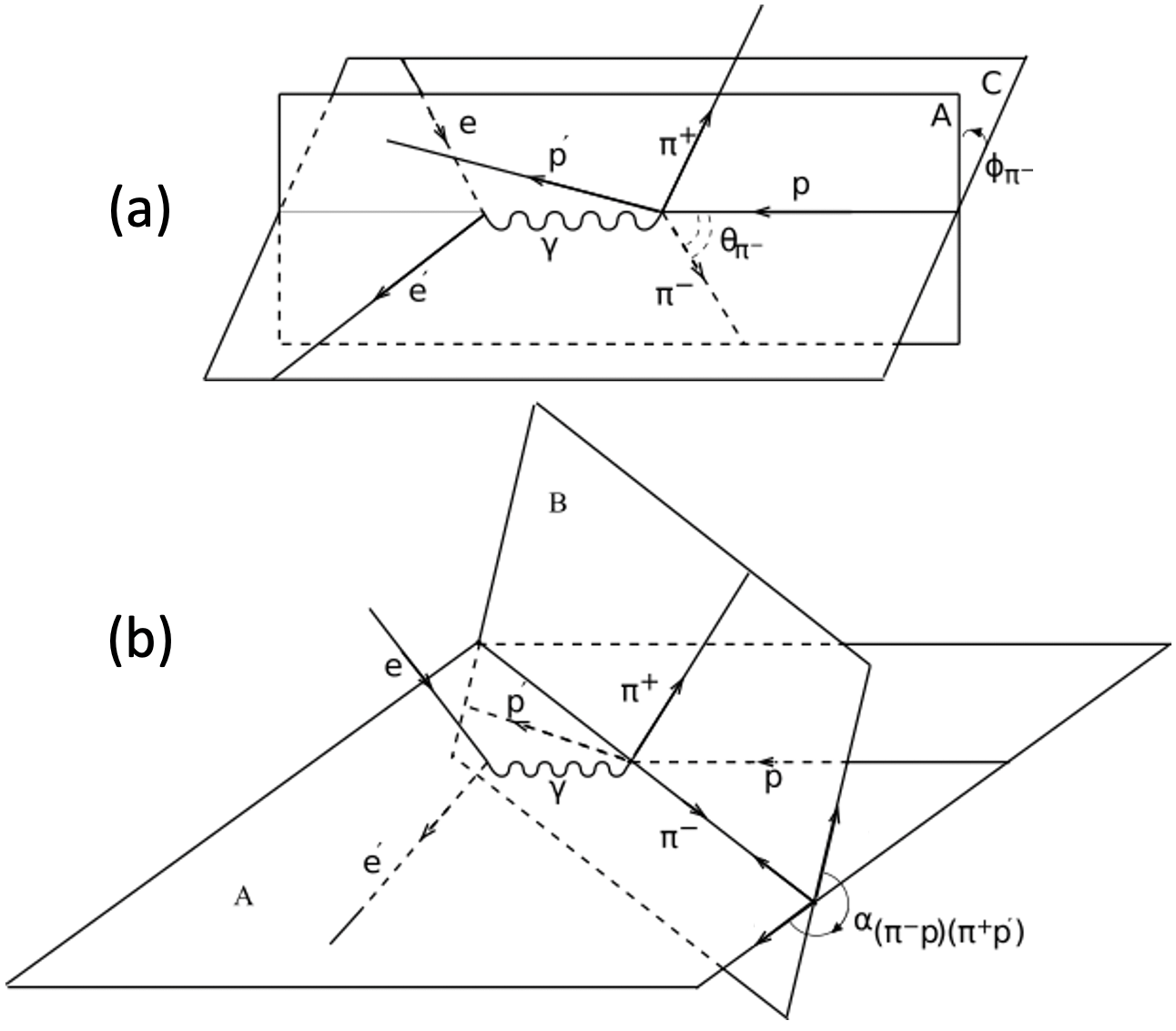}
\caption{Kinematic variables for the description of the reaction $\gamma_v p \to  \pi^+ \pi^- p'$ in the CM frame of the final-state hadrons corresponding to the $d^5\tau$ differential assignment given in Section~\ref{kinematics}. Panel (a) shows the $\pi^-$ polar and azimuthal angles $\theta_{\pi^-}$ and $\phi_{\pi^-}$. Plane C represents the electron scattering plane. The $z$-axis is directed along the $\gamma_v$ three-momentum, while the $x$-axis is located in the electron scattering plane C and the $y$-axis forms a right-handed coordinate system. Plane A is defined by the three-momenta of the initial-state proton and the final-state $\pi^-$. Panel (b) shows the angle $\alpha_{[\pi^-p][\pi^+p']}$ between the two hadronic planes A and B or the plane B rotation angle around the axis aligned along the three-momentum of the final-state $\pi^-$. Plane B is defined by the three-momenta of the final-state $\pi^+$ and $p'$.} 
\label{fig_kinematic}
\end{center}
\end{figure}

\subsection{$\pi^+\pi^-p$ Virtual Photon Cross Sections}
\label{kinematics}

In this paper, we have fit the virtual photon $\gamma_v p \to \pi^+\pi^-p'$ cross sections obtained from electron scattering assuming the single-photon exchange approximation. In this process, the invariant mass of the final-state hadrons $W$ and the photon virtuality $Q^2$ unambiguously determine the initial-state virtual photon and proton four-momenta in their center-of-mass (CM) frame, with the $z$-axis directed along the three-momentum of the virtual photon, as illustrated in Fig.~\ref{fig_kinematic}. The final $\pi^+\pi^-p$ state is described by twelve variables, corresponding to the four-momenta of the three final-state hadrons. Energy-momentum conservation and the on-shell conditions for the final-state hadrons reduce the number of independent variables to five. Thus, for a given $W$ and $Q^2$, the reaction is fully described by the five-fold differential cross section $d^5\sigma/d^5\tau$, where $d^5\tau$ denotes the differential in the five independent kinematic variables that define the final-state hadron four-momenta. There are multiple possible choices for these five variables~\cite{Byckling:1971vca}.

Defining $M_{\pi^+p'}$, $M_{\pi^-p'}$, and $M_{\pi^+\pi^-}$ as the invariant masses of the three possible two-hadron subsystems in the final state, we adopt the following choice for evaluating the five-fold differential cross section

\begin{equation}
d^5\tau = dM_{\pi^+p'} \, dM_{\pi^+\pi^-} \, d\Omega_{\pi^-} \, d\alpha_{[\pi^-p][\pi^+p']},
\end{equation}
\noindent
where $\Omega_{\pi^-}$ is the solid angle of the $\pi^-$ in the final-state CM frame, specified by the polar angle $\theta_{\pi^-}$ and azimuthal angle $\phi_{\pi^-}$ as shown in Fig.~\ref{fig_kinematic}(a). The variable $\alpha_{[\pi^-p][\pi^+p']}$ is the angle between two planes: one defined by the momenta of the $\pi^-$ and the target proton $p$, and the other by the momenta of the $\pi^+$ and the final-state proton $p'$, measured around the axis defined by the momentum of the $\pi^-$, as shown in Fig.~\ref{fig_kinematic}(b).

This choice of $d^5\tau$ is used to compute the $\pi^+\pi^-p$ differential cross sections within the JM model~\cite{Ripani:2000va, Mokeev:2008iw, Mokeev:2012vsa, Mokeev:2015lda, Mokeev:2023zhq} for comparison with the experimental data, as in our previous analyses~\cite{Mokeev:2008iw, Mokeev:2012vsa, Mokeev:2015lda, Mokeev:2023zhq}. All frame-dependent kinematic variables are defined in the CM frame of the final-state hadrons. Thus far, only unpolarized or $\phi$-independent differential cross sections have been analyzed.

\begin{table}
\begin{center}
\begin{tabular}{|c|c|} \hline
                          & $2.0-2.4$ \\ \cline{2-2}
                          & $2.4-3.0$ \\ \cline{2-2}
$Q^2$-bins (GeV$^2$)      & $3.0-3.5$ \\ \cline{2-2}
                          & $3.5-4.2$ \\ \cline{2-2}
                          & $4.2-5.0$ \\ \hline
$W$ interval (GeV)        & $1.56-1.76$ \\
covered in each $Q^2$ bin & 9 25-MeV wide bins \\ \hline
\end{tabular}
\caption{Kinematic regions covered by the CLAS $\pi^+\pi^-p$ electroproduction cross sections~\cite{Trivedi:2026fsd} where the resonant/non-resonant parameters of JM23 \cite{Mokeev:2023zhq} were adjusted to the data before the extraction of the resonance parameters from the data fit.}
\label{wq2bins1} 
\end{center}
\end{table}

\begin{table}
\setlength{\tabcolsep}{6pt} 
\renewcommand{\arraystretch}{1.2} 
\begin{center}
\begin{tabular}{|c|c|c|} \hline
One-Fold Differential                                  & Interval                                                & Number of \\
Cross Section                                          & Covered                                                 & Bins \\ \hline
$\frac{d\sigma}{dM_{\pi^+p'}}$ ($\mu$b/GeV)            & $M_{\pi^+p'}^\text{min}$-$M_{\pi^+p'}^\text{max}$       & 14 \\
$\frac{d\sigma}{dM_{\pi^+\pi^-}}$ ($\mu$b/GeV)         & $M_{\pi^+\pi^-}^\text{min}$-$M_{\pi^+\pi^-}^\text{max}$ & 14 \\
$\frac{d\sigma}{dM_{\pi^-p'}}$ ($\mu$b/GeV)            & $M_{\pi^-p'}^\text{min}$-$M_{\pi^-p'}^\text{max}$       & 14 \\
$\frac{d\sigma}{d(-\cos \theta_{\pi^-})}$ ($\mu$b/rad) & 0-180$^\circ$                                           & 10 \\
$\frac{d\sigma}{d(-\cos \theta_{\pi^+})}$ ($\mu$b/rad) & 0-180$^\circ$                                           & 10 \\
$\frac{d\sigma}{d(-\cos \theta_{p'})}$ ($\mu$b/rad)    & 0-180$^\circ$                                           & 10 \\
$d\sigma/d\alpha_{[\pi^-p][\pi^+p']}$ ($\mu$b/rad)     & 0-360$^\circ$                                           & 10 \\
$d\sigma/d\alpha_{[\pi^+p][\pi^-p']}$ ($\mu$b/rad)     & 0-360$^\circ$                                           & 10 \\
$d\sigma/d\alpha_{[\pi^+\pi^-][p p']}$ ($\mu$b/rad)    & 0-360$^\circ$                                           & 10 \\ \hline
\end{tabular}
\caption{List of the one-fold differential cross sections measured with CLAS~\cite{Trivedi:2026fsd} and the binning over the kinematic variables. $M_{i,j}^\text{min}=M_i+M_j$ and $M_{i,j}^\text{max} = W-M_k$, where $M_{i,j}$ and $M_k$ are the invariant masses of the final-state hadron pair $(i,j)$, and the mass of the third final-state hadron $k$, respectively.}
\label{1diffbins}
\end{center}
\end{table}

The $\pi^+\pi^-p$ electroproduction data have been collected in bins spanning a seven-dimensional (7D) kinematic space, since the two-dimensional phase space for the scattered electron must be included in the measured exclusive electron scattering cross sections. The number of bins in this 7D reaction phase space, along with the kinematic coverage for the extraction of the differential cross sections, is detailed in Tables~\ref{wq2bins1} and \ref{1diffbins}. The extremely large number of bins in the 7D space--approximately 10$^7$--precludes the use of fully correlated multi-fold differential cross sections in the data analysis.

Moreover, due to statistical limitations, more than half of the five-dimensional (5D) phase-space bins for the final-state hadrons remain unpopulated at any given $W$ and $Q^2$. Therefore, we restrict the analysis to the use of one-fold differential cross sections in each $(W, Q^2)$ bin covered by the data. These cross sections include

\begin{itemize}
\item invariant mass distributions for the three pairs of final-state particles $d\sigma/dM_{\pi^+\pi^-} $, $d\sigma/dM_{\pi^+p'}$, and $d\sigma/dM_{\pi^- p'}$;
\item distributions over CM polar angles of the three final-state particles $d\sigma/d(-\cos \theta_{\pi^-})$, $d\sigma/d(-\cos \theta_{\pi^+})$, and $d\sigma/d(-\cos \theta_{p'})$;
\item distributions over the three $\alpha$-angles determined in the CM frame: $d\sigma/d\alpha_{[\pi^-p][\pi^+p']}$,
$d\sigma/d\alpha_{[\pi^+p][\pi^-p']}$, and $d\sigma/d\alpha_{[\pi^+\pi^-][p p']}$, where $d\sigma/d\alpha_{[\pi^+p][\pi^-p']}$ and
$d\sigma/d\alpha_{[\pi^+\pi^-][p p']}$ are defined analogously to $d\sigma/d\alpha_{[\pi^-p][\pi^+p']}$ described above. 
\end{itemize} 

The one-fold differential cross sections were obtained by integrating the five-fold differential cross sections over the remaining four kinematic variables of $d^5\tau$. However, angular distributions involving the polar angles of the final-state $\pi^+$ and $p'$, as well as the rotation angles around the momentum directions of these hadrons, cannot be derived from $d^5\tau$ as previously defined, since this differential does not depend on those variables. To access these observables, two alternative differentials--$d^5\tau'$ and $d^5\tau''$--are required. These include $d\Omega_{\pi^+} d\alpha_{[\pi^+p][\pi^-p']}$ and $d\Omega_{p'} d\alpha_{[\pi^+\pi^-][pp']}$, respectively, as described in Refs.~\cite{Mokeev:2008iw,Mokeev:2012vsa}.

The corresponding five-fold differential cross sections over $d^5\tau'$ and $d^5\tau''$ were obtained by interpolation from the cross sections computed over $d^5\tau$. For each kinematic point in the 5D phase space defined by the variables of $d^5\tau'$ or $d^5\tau''$, the four-momenta of the three final-state hadrons were reconstructed. From these, the corresponding five kinematic variables of the $d^5\tau$ set were determined, allowing the cross section $d^5\sigma/d^5\tau$ to be interpolated at the desired kinematic point.

\subsection{Model for $\pi^+\pi^-p$ Electroproduction Cross Sections}
\label{jm_model}

For the extraction of the $\gamma_v p N^*$ electrocouplings of nucleon resonances within the mass range $W = 1.56 - 1.76$~GeV for $Q^2$ from $2.0-5.0$~GeV$^2$, the phenomenological, data-driven Jefferson Lab-Moscow State University JM reaction model \cite{Ripani:2000va, Mokeev:2008iw,Mokeev:2012vsa, Mokeev:2015lda,Mokeev:2023zhq} was employed. To date, this approach has provided all available information on the nucleon resonance electrocouplings from the $\pi^+\pi^-p$ electroproduction channel. The JM reaction model includes all relevant mechanisms needed to describe the $\pi^+\pi^-p$ electroproduction amplitudes, as manifested in the kinematic dependencies of the nine one-fold differential cross sections discussed in Section~\ref{kinematics}. The studies presented in this paper employ the most recent version of the model, JM23~\cite{Mokeev:2023zhq}.

\begin{figure*}[htp]
\begin{center}
\includegraphics[width=12.8cm]{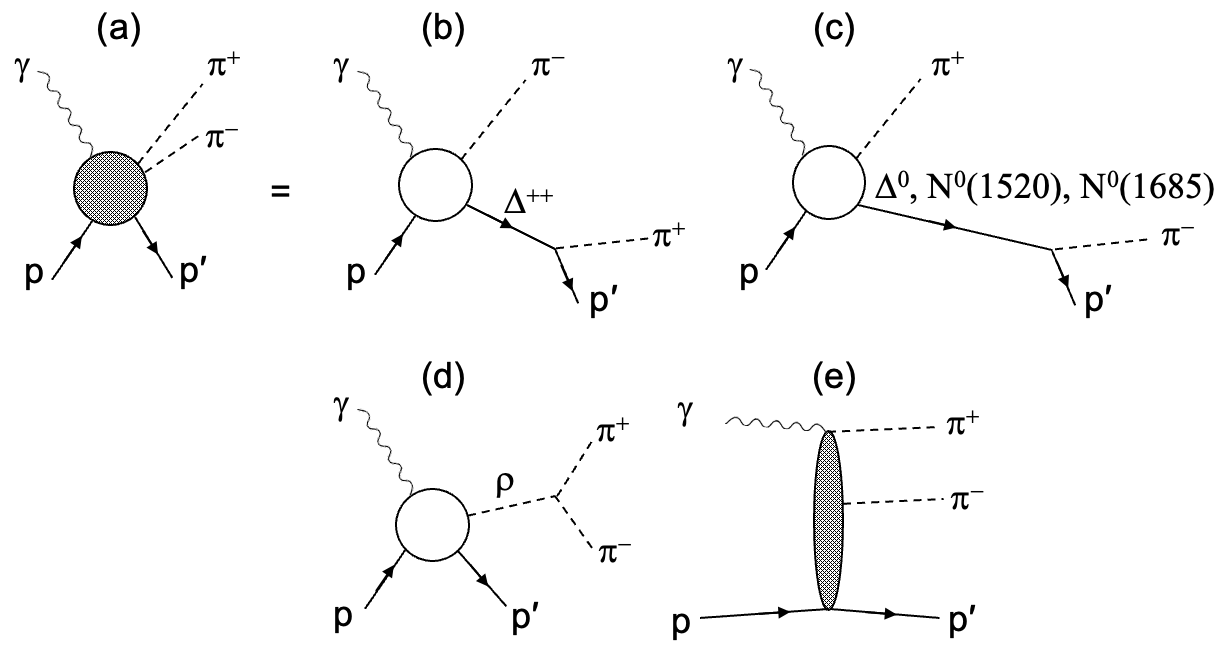}
\vspace{-3mm}
\caption{The $\gamma_v p\to  \pi^+\pi^- p'$ electroproduction mechanisms incorporated into the JM23 model \cite{Mokeev:2008iw,Mokeev:2012vsa,Mokeev:2015lda,Mokeev:2023zhq}: a) full amplitude; b) $\pi^- \Delta^{++}$; c) $\pi^+ \Delta^0$, $\pi^+ N^0(1520)3/2^-$, and $\pi^+N^0(1680)5/2^+$ subchannels; d) $\rho p$ subchannel; e) direct $2 \pi$ mechanisms.}
\label{jmmech}
\end{center}
\end{figure*}

The mechanisms incorporated into the JM model are illustrated in Fig.~\ref{jmmech}. The full amplitude of the $\gamma_v p \to \pi^+\pi^- p'$ reaction is described as a superposition of contributions from the $\pi^- \Delta^{++}$, $\pi^+ \Delta^0$, $\rho p$, $\pi^+ N^0(1520)$, and $\pi^+ N^0(1680)$ subchannels, with the unstable intermediate states decaying into the respective two-body final states, ultimately forming the three-body $\pi^+\pi^-p$ final state. The detailed description of these channels is given in Appendix III of Ref.~\cite{Mokeev:2008iw}. In addition, the model includes direct $2\pi$ production mechanisms, where the $\pi^+\pi^-p$ final state is produced without intermediate unstable hadron formation. Evidence for these direct contributions has been observed through the analysis of the final-state hadron angular distributions using phenomenological amplitudes, as discussed in Ref.~\cite{Mokeev:2008iw}. 

Within the JM23 model, only the $\pi^-\Delta^{++}$, $\pi^+\Delta^0$, and $\rho p$ subchannels include contributions from $N^*$ states excited in the $s$-channel of the $\gamma_v p$ interaction. The JM23 model incorporates contributions from all well-established $N^*$ states listed in Table~\ref{nstlist}. Studies of $\pi^+\pi^-p$ electroproduction in the third resonance region have revealed no evidence for contributions from the $N(1710)1/2^+$~\cite{Mokeev:2020hhu}. The resonance decay widths to the final hadronic states for the excited nucleon states within the mass range of $W < 1.6$~GeV were adopted from our previous analysis of $\pi^+\pi^-p$ electroproduction in Ref.~\cite{Mokeev:2023zhq}. For the excited states in the range $1.6 < W < 1.76$~GeV, the hadronic decay parameters were taken from Refs.~\cite{Mokeev:2020hhu,CLAS:2018drk}. In addition, we accounted for contributions from the tails of high-mass nucleon resonances in the mass range of $1.9 < W < 2.0$~GeV, as listed in Table~\ref{nstlist}. Since the total decay widths of these resonances may reach values as large as 350~MeV, their contributions within the third resonance region cannot be neglected. The hadronic decay parameters for these high-mass states were taken from the Particle Data Group (PDG) listings~\cite{ParticleDataGroup:2024cfk}.

The $\gamma_vpN^*$ electrocouplings for resonances within the mass range of $W < 1.6$~GeV derived from $\pi^+\pi^-p$ electroproduction data from Ref.~\cite{Mokeev:2023zhq} are consistent with the electrocouplings extracted from $\pi N$ electroproduction \cite{Aznauryan:2009mx}. As starting values for the electrocouplings of the excited proton states in the third resonance region, we employed the parameterization given in Ref.~\cite{HillerBlin:2019jgp}. These initial values were subsequently refined to achieve a good description of the experimental data on the nine one-fold $\pi^+\pi^-p$ differential cross sections. For the resonances in the mass range $1.9 < W < 2.0$~GeV, we adopted initial electrocouplings from quark model predictions~\cite{Ronniger:2012xp, Giannini:2015zia}. These values were then adjusted to describe the CLAS results on the nine one-fold differential cross sections within the $W$ interval $1.9-2.0$~GeV over the full range of $Q^2$ from 2.0 to 5.0~GeV$^2$.

\begin{table}
\begin{center}
\begin{tabular}{|c|c|c|c|} \hline
$N^*$ States        & Mass        & Total Decay               & Refs.  \\
Incorporated        &  (GeV)      & Width                     &      \\
in Data Fit         &             & $\Gamma_\text{tot}$ (GeV) &     \\ \hline
$N(1440)1/2^+$      & $1.43-1.48$ & $0.25-0.40$               & \cite{Aznauryan:2009mx} \\
$N(1520)3/2^-$      & $1.51-1.53$ & $0.12-0.13$               & \cite{Aznauryan:2009mx} \\
$N(1535)1/2^-$      & $1.51-1.55$ & $0.12-0.18$               & \cite{Aznauryan:2009mx} \\
$N(1650)1/2^-$      & $1.64-1.67$ & $0.15-0.16$               & \cite{HillerBlin:2019jgp} \\
$N(1680)5/2^+$      & $1.68-1.69$ & $0.11-0.13$               & \cite{Park:2014yea} \\
$N(1700)3/2^-$      & $1.65-1.75$ & $0.16-0.18$               & \cite{Park:2014yea} \\
$N'(1720)3/2^+$     & $1.71-1.74$ & $0.11-0.13$               & \cite{Mokeev:2020hhu} \\
$N(1720)3/2^+$      & $1.73-1.76$ & $0.11-0.13$               & \cite{HillerBlin:2019jgp} \\ 
$\Delta(1600)3/2^+$ & $1.50-1.64$ & $0.20-0.30$               & \cite{Mokeev:2023zhq} \\
$\Delta(1620)1/2^-$ & $1.60-1.66$ & $0.11-0.15$               & \cite{HillerBlin:2019jgp} \\
$\Delta(1700)3/2^-$ & $1.67-1.73$ & $0.23-0.32$               & \cite{HillerBlin:2019jgp} \\ 
$N(1900)3/2^+$      & $1.89-1.95$ & $0.10-0.32$               & \cite{Ronniger:2012xp, Giannini:2015zia} \\ 
$\Delta(1905)5/2^+$ & $1.86-1.91$ & $0.27-0.40$               & \cite{Ronniger:2012xp, Giannini:2015zia} \\ 
$\Delta(1920)3/2^+$ & $1.87-1.97$ & $0.24-0.36$               & \cite{Ronniger:2012xp, Giannini:2015zia} \\ 
$\Delta(1950)7/2^+$ & $1.92-1.95$ & $0.24-0.34$               & \cite{Ronniger:2012xp, Giannini:2015zia} \\ \hline
\end{tabular}
\caption{List of resonances included in the fit of the $\pi^+\pi^-p$ differential cross sections at $1.56 < W < 1.76$~GeV for $2.0 < Q^2 < 5.0$~GeV$^2$ and their parameters: masses, total decay widths $\Gamma_\text{tot}$, and ranges of their variation. The starting values for the resonance electrocouplings were taken from the references given in the last column.}
\label{nstlist} 
\end{center}
\end{table}

The resonant amplitudes are described using a unitarized Breit-Wigner ansatz~\cite{Mokeev:2012vsa}, which incorporates transitions between different resonances through the dressed resonance propagator. This approach ensures that the resonant amplitudes remain consistent with the constraints imposed by the general unitarity condition~\cite{Aitchison:1972ay,Kamano:2008gr}. Conservation of quantum numbers in strong interactions allows for transitions between specific pairs of $N^*$ states, namely: $N(1520)3/2^- \leftrightarrow N(1700)3/2^-$, $N(1535)1/2^- \leftrightarrow N(1650)1/2^-$, and $N(1720)3/2^+ \leftrightarrow N'(1720)3/2^+$. These transitions are incorporated into the JM23 model.

The non-resonant amplitudes in the $\pi\Delta$ subchannels are described using a minimal set of current-conserving Reggeized Born terms, as detailed in Ref.~\cite{Mokeev:2008iw}. These include the contact term, Reggeized $t$-channel $\pi$-in-flight exchange, $s$-channel nucleon exchange, and $u$-channel $\Delta$-in-flight contributions. Notably, as $W$ approaches the reaction threshold, the Reggeized $t$-channel term smoothly evolves into the $\pi$-pole term, thereby enabling the use of Reggeized Born terms even at low $W$. Initial and final-state interactions for the non-resonant contributions in the $\pi\Delta$ subchannels are taken into account following the approach developed in Ref.~\cite{Ripani:2000va}. Additionally, to avoid double counting, the contributions from the partial waves with total angular momentum $J \leq 7/2$ of the Reggeized $t$-channel amplitudes have been subtracted from the Born terms. This subtraction accounts for the overlap between Reggeized $t$-channel exchanges and $s$-channel resonance contributions--specifically those up to spin $7/2$ that are included in the current JM model version (see Table~\ref{nstlist}).

The phenomenological extra contact terms used to describe contributions to the $\pi\Delta$ amplitudes beyond the Born terms were further refined to achieve the level of agreement with the data required for the reliable extraction of the electrocouplings, as detailed in Ref.~\cite{Mokeev:2023zhq}.

The $\pi^+ N^0(1520)3/2^-$ and $\pi^+ N^0(1680)5/2^+$ subchannels are described in the JM23 model by non-resonant contributions only. The $\pi^+ N^0(1520)3/2^-$ amplitudes were derived from the non-resonant Born terms in the $\pi\Delta$ subchannels by introducing an additional $\gamma_5$ matrix to account for the opposite parities of the $\Delta(1232)3/2^+$ and $N(1520)3/2^-$ states~\cite{Mokeev:2008iw}. The magnitudes of the $\pi^+ N^0(1520)3/2^-$ amplitudes were fit independently to the data in each $(W, Q^2)$ bin. Contributions from this subchannel become relevant for $W > 1.6$~GeV.

The $\pi^+ N^0(1680)5/2^+$ contributions are observed in the data at $W > 1.65$~GeV, while they are almost negligible at lower $W$. In the JM23 model, effective $t$-channel exchange terms were used to parameterize the amplitudes of this subchannel~\cite{Mokeev:2008iw}, with their magnitudes also fitted independently in each $(W, Q^2)$ bin.

For $W > 1.65$~GeV, the non-resonant contributions to the $\rho p$ channel must be taken into account. To describe these non-resonant amplitudes within the resonance excitation region, we began with the simple ansatz developed in Ref.~\cite{Soding:1966}, which includes only $t$-channel exchanges parameterized by an exponential propagator. This approach was subsequently extended to electroproduction processes as described in Ref.~\cite{Cassel:1981sx}.

The magnitudes of the non-resonant $\rho p$ electroproduction amplitudes were fit to the nine one-fold differential cross sections in each $(W, Q^2)$ bin. We found that for squared momentum transfer between the initial and final proton $t_{pp'} < 1.0$~GeV$^2$, this simple ansatz provides a satisfactory description of the data. In the range of $t_{pp'} > 1.0$~GeV$^2$, the resonant contributions to the $\rho p$ channel exceed the non-resonant background. Therefore, this parameterization of the non-resonant amplitudes is suitable for use within the resonance excitation region.

In general, unitarity requires the inclusion of direct $2\pi$ production mechanisms in the $\pi^+\pi^- p$ electroproduction amplitudes, in which the final state is formed without proceeding through intermediate unstable hadron states~\cite{Aitchison:1978pw,Aitchison:1979ja}. These $2\pi$ production processes are distinct from the contributions of the aforementioned quasi-two-body subchannels with unstable intermediate hadrons. 

The direct $2\pi$ mechanisms are modeled as a sequence of two exchanges in the $t$- and/or $u$-channel by the set of unspecified particles. The corresponding amplitudes are parameterized as Lorentz-invariant contractions of the spin-tensors associated with the initial and final-state particles. The propagators for the exchanged particles are expressed using exponential functions. All details on the parameterization of these $2\pi$ mechanisms can be found in Refs.~\cite{Mokeev:2008iw,Mokeev:2015lda}. The magnitudes of their contributions are fit to the data independently for each $(W, Q^2)$ bin.

\begin{figure*}
\begin{center}
\includegraphics[width=1.0\textwidth]{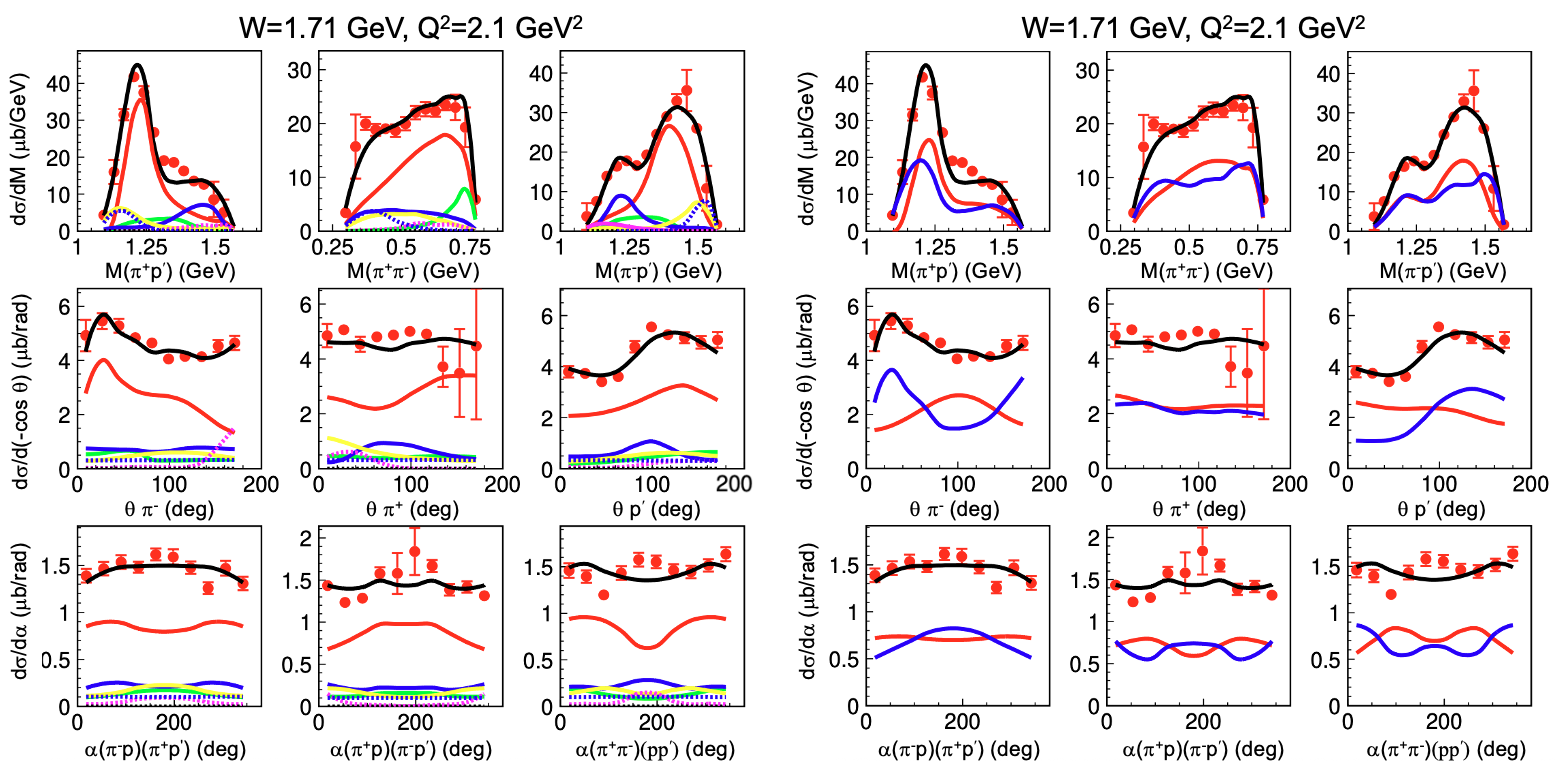}
\caption{Representative examples of the description of the nine one-fold $\pi^+\pi^-p$ differential cross sections measured with CLAS at $Q^2$=2.1~GeV$^2$~\cite{Trivedi:2026fsd} achieved within the JM23 model after initial adjustment of the resonant and non-resonant parameters to the data. (Left): Full JM23 results are shown by black solid lines. The contributions from the $\pi^-\Delta^{++}$, $\pi^+\Delta^0$, and $\rho p$ subchannels are shown by the red, blue, and green solid lines, respectively. The yellow solid, blue dotted, and magenta dotted lines correspond to the contributions from the $\pi^+N(1520)3/2^-$ and $\pi^+N(1680)5/2^+$ channels, and from direct $2\pi$ mechanisms, respectively. (Right): The red and blue lines show the  contributions from the resonant and non-resonant parts, respectively, to the full nine one-fold differential cross sections computed in the initial adjustment of the JM23 parameters (black lines). Only the statistical uncertainties are shown, except for the particular data points for the differential cross section that are either dominated by systematic uncertainties or affected by the impact from the areas of low detector efficiency or areas outside the detector acceptance.}
\label{plot_9diff1}
\end{center}
\end{figure*}

\begin{figure*}
\begin{center}
\includegraphics[width=1.0\textwidth]{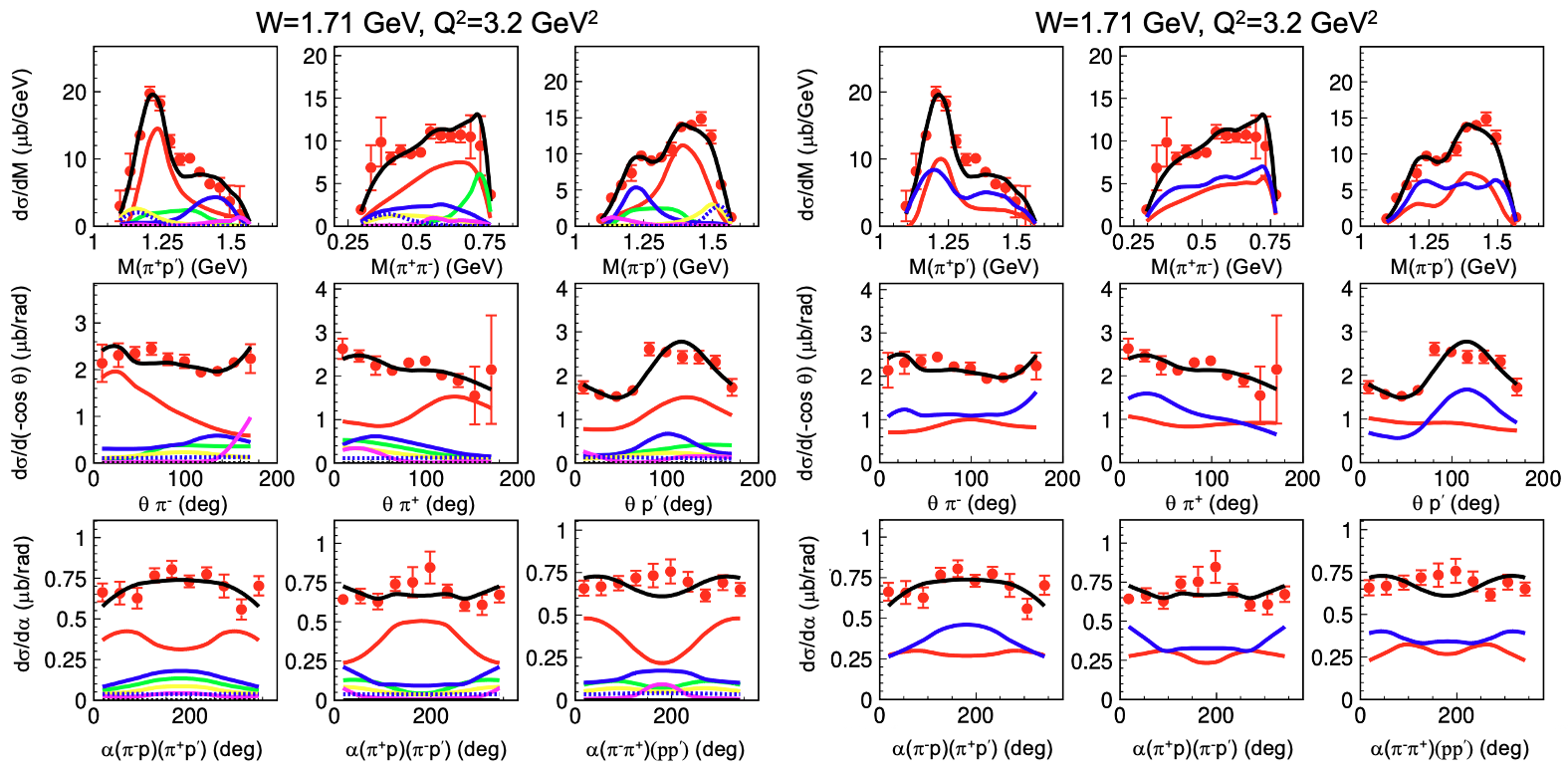}
\caption{Representative examples of the description of the nine one-fold $\pi^+\pi^-p$ differential cross sections measured with CLAS at $Q^2$=3.2~GeV$^2$ \cite{Trivedi:2026fsd} achieved within the JM23 model after initial adjustment of the resonant and non-resonant parameters to the data. See Fig.~\ref{plot_9diff1} for a description of the plots and model curves.} 
\label{plot_9diff2}
\end{center}
\end{figure*}

\begin{figure*}
\begin{center}
\includegraphics[width=1.0\textwidth]{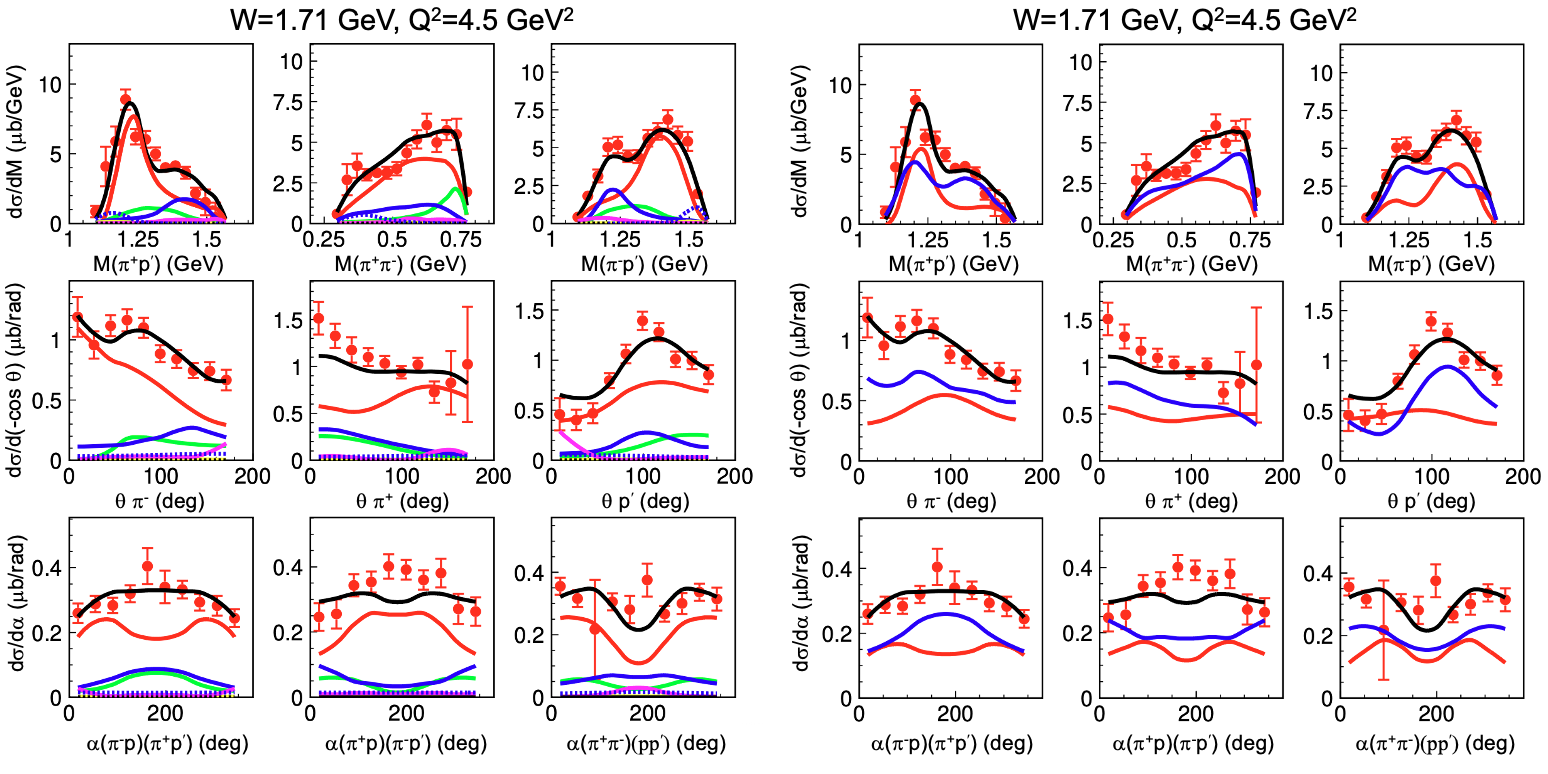}
\caption{Representative examples of the description of the nine one-fold $\pi^+\pi^-p$ differential cross sections measured with CLAS at $Q^2$=4.5~GeV$^2$ \cite{Trivedi:2026fsd} achieved within the JM23 model after initial adjustment of the resonant and non-resonant parameters to the data. See Fig.~\ref{plot_9diff1} for a description of the plots and model curves.} 
\label{plot_9diff3}
\end{center}
\end{figure*}

\begin{table*}
\begin{center}
\begin{tabular}{|c|c|c|c|c|c|} \hline
$W$ Interval (GeV) & $Q^2$: $2.0-2.4$ GeV$^2$  &$Q^2$: $2.4-3.0$ GeV$^2$ & $Q^2$: $3.0-3.5$ GeV$^2$ & $Q^2$: $3.5-4.2$ GeV$^2$ & $Q^2$: $4.2-5.0$ GeV$^2$\\  \hline
$1.58-1.60$        & 2.48      & 2.25 & 2.00 & 1.43 & 1.46 \\ \hline
$1.60-1.63$        & 2.41      & 2.64 & 1.41 & 1.71 & 1.23 \\ \hline
$1.63-1.65$        & 1.61      & 2.21 & 1.61 & 1.73 & 1.07 \\ \hline
$1.65-1.68$        & 2.62      & 2.27 & 2.20 & 1.71 & 1.60 \\ \hline
$1.68-1.70$        & 2.85      & 2.27 & 1.12 & 1.59 & 1.77 \\ \hline
$1.70-1.73$        & 2.68      & 1.78 & 1.62 & 1.54 & 1.69 \\ \hline
$1.73-1.75$        & 2.93      & 2.10 & 1.86 & 2.07 & 2.12 \\ \hline
$1.75-1.78$        & 2.77      & 2.52 & 2.05 & 2.06 & 2.60 \\ \hline
\end{tabular}
\caption{Quality of the description in terms of $\chi^2$/d.p. of the nine one-fold differential cross sections measured with CLAS~\cite{Trivedi:2026fsd} achieved after adjustment of the JM23 model \cite{Mokeev:2023zhq} parameters to the data independently in each bin of $(W,Q^2)$.}
\label{chi2dp_1diff} 
\end{center}
\end{table*}

After the initial adjustment of the resonant and non-resonant parameters of the JM23 model, a reasonable description of the nine one-fold $\pi^+\pi^-p$ differential cross sections was achieved across all bins within the kinematic range of $1.56 < W < 1.76$~GeV for $2.0 < Q^2 < 5.0$~GeV$^2$. The quality of the data description, in terms of $\chi^2$/d.p. (d.p. $\equiv$ data point), is summarized in Table~\ref{chi2dp_1diff}. The $\chi^2/\text{d.p.}$ values were evaluated from point-by-point comparisons between the measured and computed nine one-fold differential cross sections in each bin of $(W,Q^2)$. Representative examples of the data description for the $W$ bin $1.700-1.725$~GeV, near the peak of the third resonance region, are shown in Figs.~\ref{plot_9diff1}-\ref{plot_9diff3}. The plots on the left show the comparison between the measured differential cross sections and the computed contributions from the quasi-two-body meson-baryon channels involving unstable hadrons, as well as from direct $2\pi$ production mechanisms. The decomposition of the computed differential cross sections into resonant and non-resonant contributions is shown in the plots on the right in Figs.~\ref{plot_9diff1}-\ref{plot_9diff3}. No evidence was found in the data to justify the inclusion of mechanisms beyond those already incorporated in the current JM23 model version.

Since the $\chi^2/\text{d.p.}$ values in Table~\ref{chi2dp_1diff} are based primarily on statistical uncertainties, they indicate a satisfactory data description suitable for initiating the $N^*$ parameter extraction as detailed in Section~\ref{fit_strategy}. The observed decrease in $\chi^2/\text{d.p.}$ with $Q^2$ reflects the increasing experimental uncertainties, while the resonant and non-resonant contributions remain comparable across the $Q^2$ range analyzed.

The contributions to the differential cross sections shown in Figs.~\ref{plot_9diff1}-\ref{plot_9diff3} (left) from various $\pi^+\pi^-p$ electroproduction mechanisms exhibit notable differences in shape within each individual differential cross section. Moreover, the contribution from any given mechanism varies in shape across the different cross sections. These variations are governed by the underlying reaction dynamics. Such distinctive features enable us to identify the contributing mechanisms and gain insight into their amplitude parameterization by simultaneously fitting all nine one-fold differential cross sections. The results presented in Figs.~\ref{plot_9diff1}-\ref{plot_9diff3} (right) reveal that the resonant and non-resonant contributions are of comparable magnitude, yet their shapes differ across the nine cross sections. This shape differentiation enables the isolation of the resonant contributions. The $N^*$ parameters are then extracted from these resonant contributions, which are modeled using a unitarized Breit-Wigner ansatz~\cite{Mokeev:2012vsa}, ensuring consistency with the general unitarity condition.

\section{Resonance Parameter Extraction From Cross Section Fits}
\label{fit_strategy}

The electrocouplings of the $s$-channel resonances $\Delta(1600)3/2^+$, $N(1675)5/2^-$, $N(1680)5/2^+$, $\Delta(1700)3/2^-$, $N(1720)3/2^+$, and $N'(1720)3/2^+$, as well as their branching fractions for decays into the $\pi\Delta$ and $\rho p$ channels, were extracted from fits to the nine independent one-fold $\pi^+\pi^-p$ differential cross sections within the $(W,Q^2)$ regions listed in Table~\ref{Q2vsW_areas}. The data in each of these regions were analyzed independently.  Five of the six resonances listed above (except for the $\Delta(1600)3/2^+$) contribute in each of the three given $W$ intervals, although the non-resonant amplitudes differ across them. The consistent results obtained for the $\gamma_v p N^*$ electrocouplings from independent fits within the three $W$ intervals provide evidence for the reliable extraction of these quantities from the $\pi^+\pi^-p$ data.

\begin{table}
\begin{center}
\begin{tabular}{|c|c|c|c|} \hline
$Q^2$-interval    & \multicolumn{3}{c|}{$W$-intervals (GeV)} \\ \hline
$2.0-3.5$ GeV$^2$ & $1.56-1.66$              & $1.61-1.71$              & $1.66-1.76$  \\ \hline
$3.0-3.5$ GeV$^2$ & $1.56-1.66$              & $1.61-1.71$              & $1.66-1.76$  \\ \hline
\end{tabular}
\caption{The kinematic regions ($Q^2$ vs. $W$) where the nine one-fold differential cross sections were fit independently for extraction of the  $\Delta(1600)3/2^+$, $N(1675)5/2^-$, $N(1680)5/2^+$, $\Delta(1700)3/2^-$, $N'(1720)3/2^+$, and $N(1720)3/2^+$ electrocouplings, masses, and total and partial decay widths to the $\pi\Delta$ and $\rho p$ final states.}
\label{Q2vsW_areas}
\end{center}
\end{table}

In the data fits, we simultaneously varied the electrocouplings, the resonance partial decay widths into $\pi \Delta$ and $\rho p$, and the Breit-Wigner masses for all $N^*$ states listed in Table~\ref{nstlist}. For resonances with masses below 1.6~GeV, their masses and total/partial hadronic decay widths were allowed to vary within the uncertainties established in previous analyses of $\pi^+\pi^-p$ electroproduction data~\cite{Mokeev:2023zhq}. For resonances with masses above 1.6~GeV, their masses and decay widths were varied within the intervals quoted in the PDG~\cite{ParticleDataGroup:2024cfk}.

The starting values for the resonance decay amplitudes into $\pi\Delta$ and $\rho p$ of orbital angular momentum $L$ and total spin $S$ were defined as
\begin{equation}
\sqrt{\Gamma^i_\text{LS}} = \sqrt{\Gamma_\text{tot} \cdot \text{BF}^i_\text{LS}},
\label{hadr_decay}
\end{equation}
where the resonance total decay widths $\Gamma_\text{tot}$ were taken from Ref.~\cite{ParticleDataGroup:2024cfk}, and the branching fractions BF$^i_\text{LS}$ ($i = \pi\Delta, \rho p$) were taken from previous CLAS studies of $\pi N$ and $\pi^+\pi^-p$ electroproduction data~\cite{Aznauryan:2009mx,Park:2014yea,CLAS:2018drk,Mokeev:2020hhu} for resonances in the mass range up to 1.75~GeV. For high-lying resonances with masses above 1.85~GeV, the masses, total decay widths, and partial hadronic decay widths were taken from Ref.~\cite{ParticleDataGroup:2024cfk}. For each resonance, the total decay width was computed as the sum of all partial decay widths.

The starting values for the electrocouplings of the $N(1440)1/2^+$, $N(1520)3/2^-$, and $\Delta(1600)3/2^+$ were taken from Ref.~\cite{Mokeev:2023zhq}. In the data fits, the electrocouplings and decay widths of these resonances were allowed to vary within the intervals established in that previous analysis~\cite{Mokeev:2023zhq}. This procedure ensured that the contributions from the tails of resonances with masses below 1.6~GeV were properly accounted for in the $W$ interval from 1.6 to 1.76~GeV.

The starting values for the electrocouplings of the resonances listed in Table~\ref{nstlist} with masses above 1.6~GeV were taken from a preliminary adjustment of the resonant and non-resonant parameters of the JM23 model to the nine one-fold $\pi^+\pi^-p$ differential cross sections, as described in Section~\ref{jm_model}. For resonances in the range 1.9-2.0~GeV, the initial electrocoupling values were taken from the models of Refs.~\cite{Giannini:2015zia, Ronniger:2012xp}. These values were further tuned in the adjustment of the JM model parameters to the nine one-fold $\pi^+\pi^-p$ electroproduction cross sections measured in the kinematic region $1.9 < W < 2.0$~GeV and $2.0 < Q^2 < 5.0$~GeV$^2$~\cite{Trivedi:2026fsd}.

The electrocouplings obtained after this adjustment were then used as the starting values in the data fits for the region $1.56 < W < 1.76$~GeV and $2.0 < Q^2 < 5.0$~GeV$^2$, to account for the contributions from the tails of resonances with masses above 1.9~GeV. For these high-lying $N^*$ states, the electrocouplings, masses, and total/partial hadronic decay widths were kept fixed at the starting values and not varied in the fits. In the fits, we varied the electrocouplings of all $N^*$s of four-star status other than the $N(1440)1/2^+$, $N(1520)3/2^-$, $\Delta(1600)3/2^+$, within the mass range from 1.6~GeV to 1.75~GeV. This variation employed a normal distribution with the $\sigma$ width parameters equal to 20\% of their starting values described above. The variations of the resonance masses and total widths $\Gamma_\text{tot}$, induced by changes in the partial hadronic decay widths into $\pi\Delta$ and $\rho p$, were constrained to remain within the intervals reported in Ref.~\cite{ParticleDataGroup:2024cfk}. In this way, restrictions were imposed on the allowed variations of the $N^*$ partial hadronic decay widths $\Gamma^i_\text{LS}$ ($i = \pi\Delta, \rho p$), decomposed over the $LS$ partial waves.

In the data fits, we also varied the following parameters of the non-resonant mechanisms included in the JM23 model
\begin{itemize}
\item the magnitudes of the additional contact-term amplitudes in the $\pi^- \Delta^{++}$ and $\pi^+ \Delta^0$ channels (one parameter per $Q^2$ bin);
\item the magnitudes of the $\pi^+ N(1520)3/2^-$ channel (one parameter per $Q^2$ bin);
\item the magnitudes of all direct $2\pi$ production amplitudes (up to six parameters per $Q^2$ bin).
\end{itemize}

The starting values for these parameters were obtained from the initial adjustment to the nine independent one-fold $\pi^+\pi^-p$ differential cross sections. To preserve the $W$ dependence established in that adjustment, we introduced $W$-independent multiplicative factors applied to the magnitudes of the non-resonant amplitudes listed above. These factors were kept fixed across the entire $W$ interval for a given $Q^2$ bin, but were varied independently between bins. The multiplicative factors were allowed to fluctuate around unity, following normal distributions with $\sigma$ values of about 20\%. This procedure maintained the smooth $W$ dependence of the non-resonant contributions, while enabling improvements in the description of the data through the simultaneous variation of both the resonant and non-resonant parameters.

For each trial set of JM23 resonant and non-resonant parameters, we computed the nine one-fold $\pi^+\pi^-p$ differential cross sections and the corresponding $\chi^2/\text{d.p.}$ values. The latter were obtained from point-by-point comparisons between the measured and computed cross sections within the $(W,Q^2)$ bins located in the regions listed in Table~\ref{chidp}. The nine one-fold differential cross sections in each of these six kinematic regions were fit independently.

The data uncertainties include both statistical uncertainties and the kinematically dependent part of the systematic uncertainties, combined in quadrature. In the fits, we selected the computed cross sections that were closest to the data and satisfied the condition $\chi^2/\text{d.p.}$ $< \chi^2_{\text{max}}/\text{d.p.}$. The values of $\chi^2_{\text{max}}/\text{d.p.}$ were chosen such that the computed cross sections remained within the data uncertainties for the majority of data points. This procedure yielded $\chi^2/\text{d.p.}$ intervals within which the computed cross sections describe the data equally well within uncertainties.

\begin{table}
\begin{center}
\begin{tabular}{|c|c|c|c|} \hline
\multirow{2}{*}{$Q^2$ Interval} & \multicolumn{3}{c|}{$W$ Interval (GeV)} \\ \cline{2-4}
                                & $1.56-1.66$ & $1.61-1.71$ & $1.66-1.76$ \\ \hline
$2.0 < Q^2 < 3.5$~GeV$^2$       & $0.77-0.81$ & $0.87-0.91$ & $0.85-0.93$ \\ \hline
$3.0 < Q^2 < 5.0$~GeV$^2$       & $0.82-0.89$ & $0.84-0.94$ & $0.92-0.98$ \\ \hline
\end{tabular}
\caption{The ranges of $\chi^2$/d.p. for the nine one-fold $\pi^+\pi^-p$ differential cross sections selected in the data fit computed within JM23 in overlapping $Q^2$ vs. $W$ intervals. The uncertainties for the measured data are given by the quadrature sum of the statistical and that part of the systematic uncertainty dependent on the final-state hadron kinematics.}
\label{chidp} 
\end{center}
\end{table}

\begin{figure*}[htbp]
\begin{center}
\includegraphics[width=1.0\textwidth]{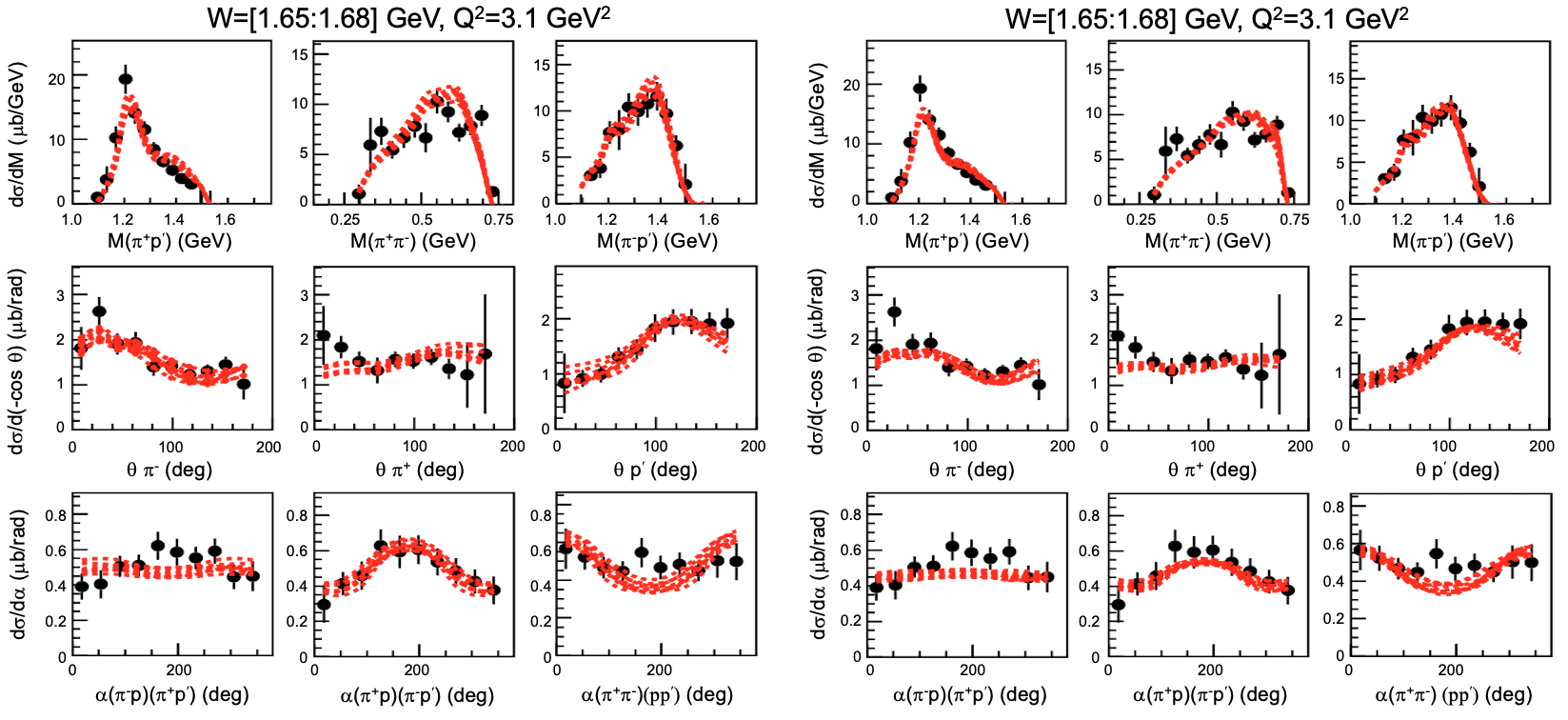}
\caption{Representative examples for the computed one-fold $\pi^+\pi^-p$ differential cross sections (groups of red curves) selected in the data fit in comparison with the measurements from CLAS~\cite{Trivedi:2026fsd} in the $W$-bin $[1.65:1.68]$~GeV. The left and right collection of $3 \times 3$ plots show the independent fits within the $W$ intervals $[1.56:1.66]$~GeV and $[1.61:1.71]$~GeV, respectively, for the $Q^2$-interval $[3.0:3.5]$~GeV$^2$.} 
\label{dcs9-w1.61}
\end{center}
\end{figure*}

\begin{figure*}[htbp]
\begin{center}
\includegraphics[width=1.0\textwidth]{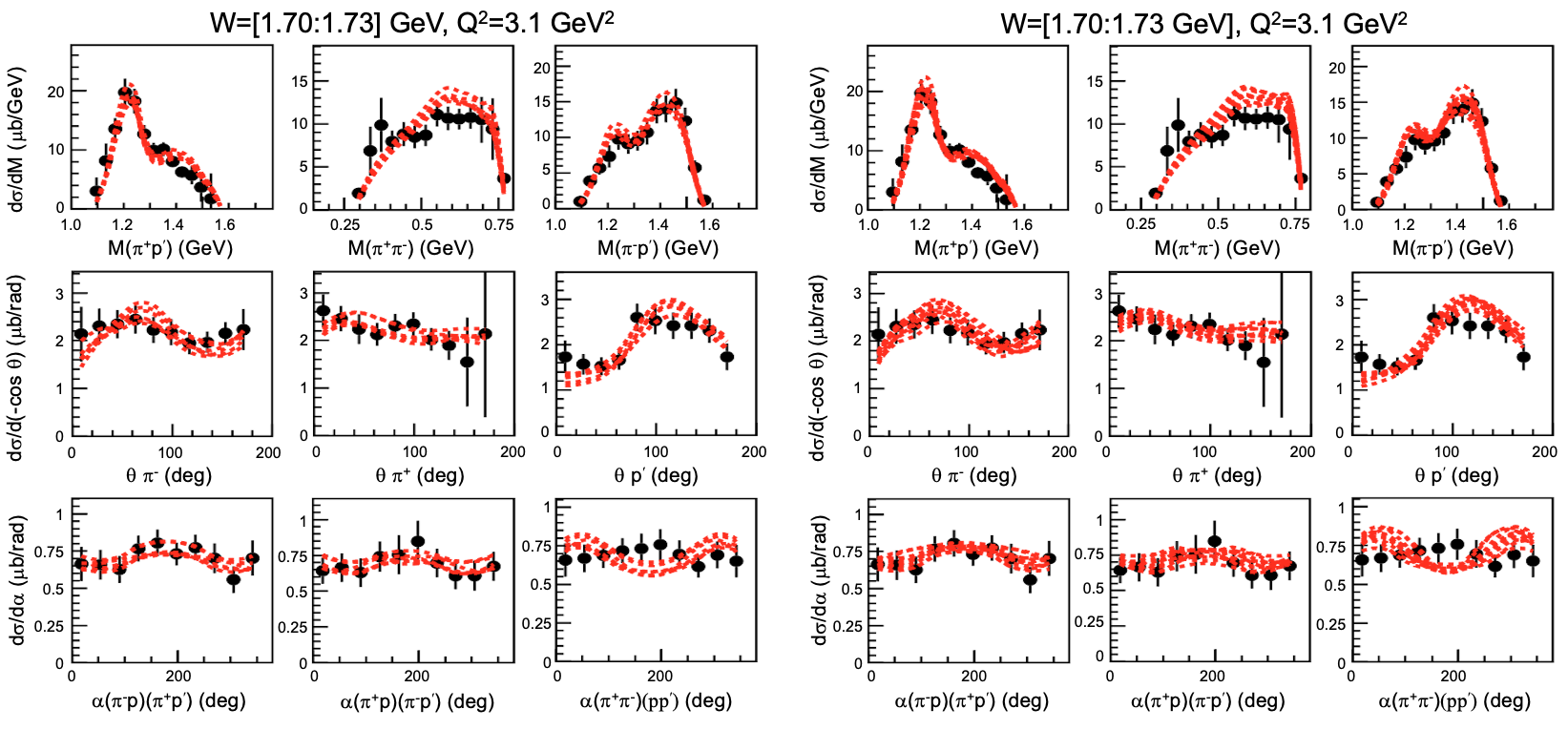}
\caption{Representative examples for the computed one-fold $\pi^+\pi^-p$ differential cross sections (groups of red curves) selected in the data fit in comparison with the measurements from CLAS~\cite{Trivedi:2026fsd} in the $W$-bin $[1.70:1.73]$~GeV. The left and right collection of $3 \times 3$ plots show the independent fits within the $W$ intervals $[1.61:1.71]$~GeV and $[1.66:1.76]$~GeV, respectively, for the $Q^2$-interval $[3.0:3.5]$~GeV$^2$.} 
\label{dcs9-w1.71}
\end{center}
\end{figure*}

For the computed cross sections selected from the data fits, the $\chi^2/\text{d.p.}$ values remain below 1.0 across the entire $(W,Q^2)$ range studied. This confirms that the computed cross sections chosen as closest to the data remain within the experimental uncertainties for most measured data points. Moreover, the relatively narrow $\chi^2/\text{d.p.}$ intervals listed in Table~\ref{chidp} demonstrate that all cross sections selected in the fits describe the data with comparable quality. Representative examples of the computed differential cross sections selected from the data fits, compared with the measured results~\cite{Trivedi:2026fsd}, are shown in Figs.~\ref{dcs9-w1.61} and \ref{dcs9-w1.71}. These figures present fit results obtained independently within overlapping $W$ intervals. The same nucleon resonances contribute across these intervals, while the non-resonant contributions differ. The good and consistent descriptions of the data suggest a credible separation between the resonant and non-resonant contributions. The resonance parameters were extracted from their evaluated contributions by fitting them with the unitarized Breit-Wigner ansatz~\cite{Mokeev:2012vsa}, while enforcing the constraints imposed by the general unitarity condition for the resonant amplitudes.

\section{Resonance Electrocouplings and Decay Widths}
\label{elcoupl_hadrdec}

The resonance parameters determined from the data fit include the electrocouplings, the total decay widths, and the partial decay widths into $\pi\Delta$ and $\rho p$. They were averaged from the group of fits selected within the $\chi^2/\text{d.p.}$ ranges for the $Q^2$ vs. $W$ intervals listed in Table~\ref{chidp}. Their mean values were taken as the resonance parameters extracted from the data, and the root mean square (RMS) dispersions in these parameters were taken as the uncertainties. The electrocoupling uncertainties obtained in this manner take into account both the statistical and systematic uncertainties in the data, as well as the uncertainties associated with the JM23 model parameters for description of both the resonant and non-resonant contributions. Furthermore, we consistently account for the correlations between the variations of the resonant and non-resonant contributions when extracting the resonance parameters. In the cases where the ranges of the extracted resonance parameters (both electrocouplings and decay widths to $\pi \Delta$ and $\rho p$) covered more than 90\% of the intervals for their variation (starting values $\pm \sigma$) employed in the data fit, we further increased the ranges of the variation and repeated the data fit, so that eventually the electrocouplings and the extracted decay widths were within the intervals of the variations of the data fits. In this way, we made sure that the employed ranges were sufficient to determine both the mean values of the resonance parameters and their uncertainties. 

A special procedure was developed for the evaluation of the transverse $A_i$ ($i = 1/2$, $3/2$) and longitudinal $S_i$ ($i = 1/2$) electrocouplings analyzing the results from independent fits of the electroproduction cross sections in the three $W_j$ intervals ($j$=1,2,3) with electrocoupling central values and uncertainties $A_{i,j} \pm \delta A_{i,j}$ and $S_{i,j} \pm \delta S_{i,j}$. 
First, we found the overlap range for the electrocouplings [$A_i^\text{min}-A_i^\text{max}$] ($i = 1/2$, $3/2$) and [$S_i^\text{min}-S_i^\text{max}$] ($i = 1/2$) from the data fit in the three $W_j$ intervals
\begin{eqnarray}
\label{el_ranges}
A_i^\text{max} = \text{min}[A_{i,j}+\delta A_{i,j}] \\
S_i^\text{max} = \text{min}[S_{i,j}+\delta S_{i,j}] \nonumber \\
A_i^\text{min} = \text{max}[A_{i,j}-\delta A_{i,j}] \nonumber \\
S_i^\text{min} = \text{max}[S_{i,j}-\delta S_{i,j}]. \nonumber
\end{eqnarray}
Within these ranges, the best data description was achieved in the three overlapping $W_j$ intervals. Consequently, the central values for $A_i$ and $S_i$ were redefined as
\begin{eqnarray}
\label{el_mean}
A_i=\frac{A_i^\text{min}+A_i^\text{max}}{2},~ (i=1/2,3/2) \\ 
S_i=\frac{S_i^\text{min}+S_i^\text{max}}{2},~ (i=1/2).\nonumber
\end{eqnarray}

There are three sources of uncertainty in the evaluation of $A_i$ and $S_i$: a) the range of overlap between the electrocouplings determined from the data fit defined in Eq.(\ref{el_ranges}), b) the RMS for the spread of the central values of the determined electrocouplings in each $W_j$ ($j$=1,2,3) interval in independent fits, {\it i.e.} RMS [$A_{i,j}$] and RMS [$S_{i,j}$], and c) the differences between the central values of the redefined electrocouplings $A_i$, $S_i$ according to Eq.(\ref{el_mean}) and after averaging of the central values $A_{i,j}$, $S_{i,j}$ determined from the data fits across the three $W_j$ intervals (see $\Delta A_i$ and $\Delta S_{i}$ in Eq.~(\ref{uncertainties}). The total uncertainties were obtained as the quadrature sum of these contributions
\begin{eqnarray}
\label{uncertainties}
\delta A_i=\sqrt{\frac{(A_i^\text{max}-A_i^\text{min})^2}{4}+(\text{RMS}[A_{i,j}])^2+\Delta A_i^2} \, \\
\Delta A_i=\frac{A_i^\text{min}+A_i^\text{max}}{2}-\frac{\sum_{j=1,2,3} A_{i,j}}{3} \nonumber \\ [0.5ex]
\delta S_i=\sqrt{\frac{(S_i^\text{max}-S_i^\text{min})^2}{4}+(\text{RMS}[S_{i,j}])^2+\Delta S_i^2}\, \nonumber \\ 
\Delta S_i=\frac{S_i^\text{min}+S_i^\text{max}}{2}-\frac{\sum_{j=1,2,3} S_{i,j}}{3}. \nonumber
\end{eqnarray} 
We consider these results as the final electrocouplings from the analysis of the $\pi^+\pi^-p$ electroproduction cross sections within the JM23 model.

In the following sections we present the results for the electrocouplings and decay widths for the $\Delta(1600)3/2^+$, $N(1675)5/2^-$, $N(1680)5/2^+$, $\Delta(1700)3/2^-$, $N(1720)3/2^+$, and $N'(1720)3/2^+$ obtained from analyses of the CLAS $\pi^+\pi^-p$ electroproduction data \cite{Trivedi:2026fsd} within the kinematics areas listed in Table~\ref{chidp}.

\subsection{$\Delta(1600)3/2^+$ Resonance}
\label{delta1600}

The first results on the electrocouplings of the $\Delta(1600)3/2^+$ were obtained from the analysis of CLAS $\pi^+\pi^-p$ electroproduction data~\cite{Trivedi:2026fsd} within the JM23 reaction model and published in Ref.~\cite{Mokeev:2023zhq}. The $\Delta(1600)3/2^+$ parameters were extracted from data in the $W$ range $1.46-1.66$~GeV for $2.0 < Q^2 < 5.0$~GeV$^2$. In that analysis, the electrocouplings and decay widths of nucleon resonances with masses above 1.6~GeV were taken from Ref.~\cite{HillerBlin:2019jgp}, after adjustment to the nine one-fold differential cross sections at $W>1.6$~GeV and $2.0 < Q^2 < 5.0$~GeV$^2$. These parameters were then kept fixed in the fit of the $\pi^+\pi^-p$ cross sections in the $1.46-1.66$~GeV interval in Ref.~\cite{Mokeev:2023zhq}.

In this study, we investigated the impact of the tails of nucleon resonances in the range $1.60-1.76$~GeV on the extracted parameters of the $\Delta(1600)3/2^+$, which were determined from fits to the nine one-fold $\pi^+\pi^-p$ differential cross sections in two overlapping $W$ intervals, $1.56-1.66$~GeV and $1.61-1.71$~GeV, for two overlapping $Q^2$ intervals, $2.0-3.5$~GeV$^2$ and $3.0-5.0$~GeV$^2$. The parameters of the nucleon resonances in the $1.60-1.76$~GeV mass range were varied and fit to the data as described in Section~\ref{fit_strategy}, while those of the $\Delta(1600)3/2^+$ were varied within the uncertainties established in our previous analysis~\cite{Mokeev:2023zhq} for $W < 1.66$~GeV and $Q^2 = [2.0-5.0]$~GeV$^2$.

In the present work, the quality of the data description in the $1.56-1.66$~GeV interval is reasonable and comparable to that obtained in the previous analysis~\cite{Mokeev:2023zhq}. In the $1.61-1.71$~GeV interval, the computed cross sections from the selected fits also reproduce the data well, with $\chi^2/\text{d.p.} < 0.94$. In extracting the resonance parameters from the fits, the experimental data uncertainties were taken as the quadratic sum of the statistical uncertainties and the point-to-point systematic uncertainties.

The mass, total decay width ($\Gamma_\text{tot}$), partial decay width to the $\pi\Delta$ final state ($\Gamma_{\pi\Delta}$), and the corresponding branching fraction (BF$_{\pi\Delta}$) for the $\Delta(1600)3/2^+$, as determined from the data fit, are summarized in Table~\ref{hadr_delta1600}. The values for the mass and decay widths obtained in the present analysis are consistent with those from our previous studies, agree within the two overlapping $W$ ranges, and are compatible with the intervals reported by the PDG.

\begin{table*}
\begin{center}
\begin{tabular}{|c|c|c|c|c|c|} \hline
                                               & $\Gamma_\text{tot}$ (MeV) & $\Gamma_{\pi\Delta}$ (MeV) & BF($\pi\Delta$) (\%)   & Mass (GeV)       & Ref. \\ \hline
$W$: $1.56-1.66$ GeV                           & $234 \pm 22$              & $144 \pm 22$               & $48-78$               & $1.57 \pm 0.023$ & current analysis\\
$Q^2$: $2.0-3.5$ GeV$^2$                       & $256 \pm 33$              & $166 \pm 34$               & $46-90$               & $1.57 \pm 0.018$ & \cite{Mokeev:2023zhq} \\ \hline \hline
$W$: $1.56-1.66$ GeV                           & $239 \pm 27$              & $149 \pm 27$               & $47-79$               & $1.58 \pm 0.031$ & current analysis \\
$Q^2$: $3.0-5.0$ GeV$^2$                       & $263 \pm 29$              & $172 \pm 29$               & $49-86$               & $1.58 \pm 0.039$ & \cite{Mokeev:2023zhq} \\ \hline
$W$: $1.61-1.71$ GeV, $Q^2$: $2.0-3.5$ GeV$^2$ & $244 \pm 35$              & $154 \pm 35$               & $43-90$               & $1.57 \pm 0.027$ & current analysis \\ \hline
$W$: $1.61-1.71$ GeV, $Q^2$: $3.0-5.0$ GeV$^2$ & $239 \pm 17$              & $149 \pm 17$               & $51-75$               & $1.59 \pm 0.032$ & current analysis \\ \hline
PDG                                            & $200-300$                 &                            & $58-82$               & $1.50-1.64$      & \cite{ParticleDataGroup:2024cfk} \\ \hline
\end{tabular}
\caption{The mass, total decay width ($\Gamma_\text{tot}$), partial $\pi\Delta$ ($\Gamma_{\pi\Delta}$) decay width, and branching fraction (BF$_{\pi\Delta}$) for the $\Delta(1600)3/2^+$ determined from the $\pi^+\pi^-p$ differential cross section fits. For the $W$ interval $1.56-1.66$~GeV, both the results from the current analysis and the previous analysis~\cite{Mokeev:2023zhq} are shown. For the $W$ interval $1.61-1.71$~GeV, only the results from the current analysis are shown.}
\label{hadr_delta1600} 
\end{center}
\end{table*}

According to the results in Table~\ref{hadr_delta1600}, there is no evidence for an evolution of the $\Delta(1600)3/2^+$ Breit-Wigner mass, total decay width, or partial decay widths to the $\pi\Delta$ and $\rho p$ final states with $Q^2$ over the broad range $2.0-5.0$~GeV$^2$. This stability suggests that the $\Delta(1600)3/2^+$ is an $s$-channel resonance excited in virtual photon-proton interactions, containing an inner core of three dressed quarks. These findings are supported by evaluations within CSMs~\cite{Liu:2022ndb,Lu:2019bjs, Cheng:2025sdp}, but they stand in tension with the claims of Ref.~\cite{Hockley:2024ipz}, which argues that the $\Delta(1600)3/2^+$ arises from strong rescattering in the $\pi N$ and $\pi\Delta$ channels. To further clarify the role of meson-baryon cloud contributions to the structure of the $\Delta(1600)3/2^+$, results on its parameters at $Q^2 < 1.0$~GeV$^2$ are essential. The CLAS $\pi^+\pi^-p$ electroproduction data~\cite{CLAS:2018fon} provide a promising opportunity to extract the $\Delta(1600)3/2^+$ parameters in this low-$Q^2$ region.

\begin{figure*}
\begin{center}
\includegraphics[width=0.9\textwidth]{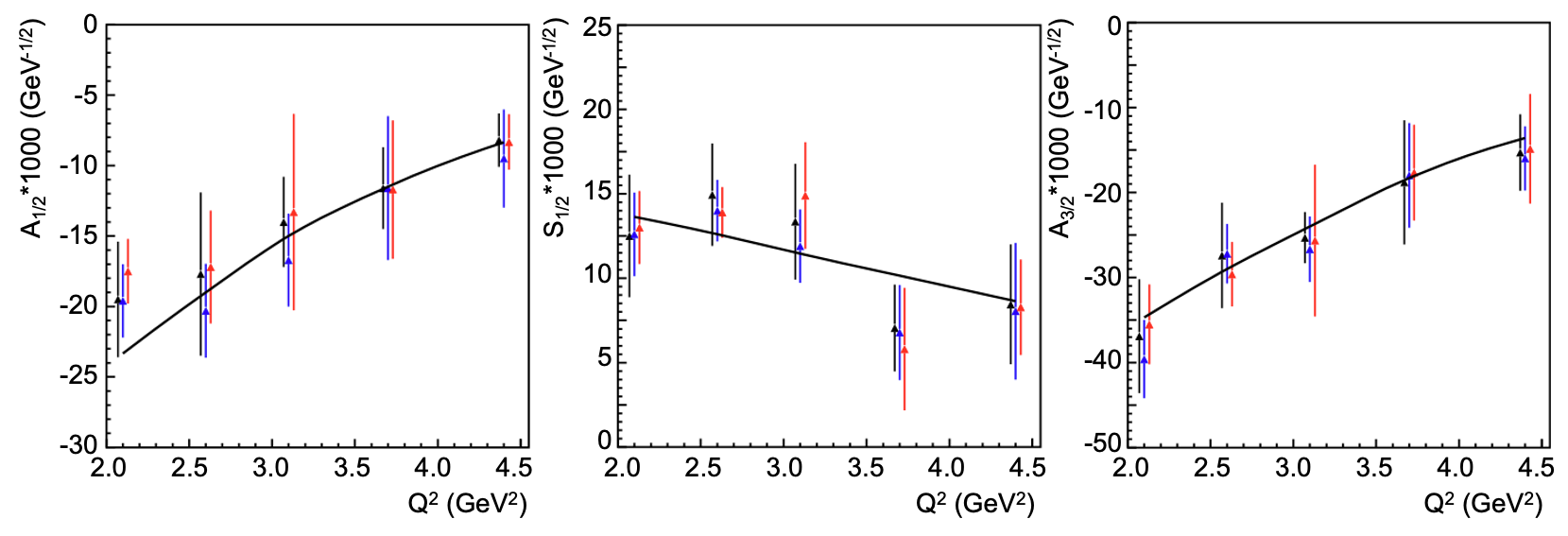}
\caption{Electrocouplings of the $\Delta(1600)3/2^+$, $A_{1/2}$ (left), $S_{1/2}$ (center), and $A_{3/2}$ (right), determined in the current independent analyses across two $W$ intervals $1.56-1.66$~GeV (blue) and $1.61-1.71$~GeV (red) in comparison with the results from the previous analysis (black) \cite{Mokeev:2023zhq} within the $W$ interval $1.56-1.66$~GeV. The CSM predictions~\cite{Lu:2019bjs} are shown by black solid lines.}  
\label{d33_1600_156166161171}
\end{center}
\end{figure*}

The $\Delta(1600)3/2^+$ electrocouplings extracted from independent fits to the data in the $W$ intervals $1.56-1.66$~GeV and $1.61-1.71$~GeV are shown in Fig.~\ref{d33_1600_156166161171}, together with the results from the previous analysis of $\pi^+\pi^-p$ electroproduction data reported in Ref.~\cite{Mokeev:2023zhq}. The present study demonstrates that the influence of the tails of nucleon resonances from the third resonance region--whose parameters were determined in fits to $\pi^+\pi^-p$ electroproduction data~\cite{CLAS:2017fja,Trivedi:2026fsd}--does not shift the $\Delta(1600)3/2^+$ parameters for either hadronic decays or electrocouplings beyond the ranges established in the earlier analysis~\cite{Mokeev:2023zhq}.

Predictions for the $\Delta(1600)3/2^+$ electrocouplings made in 2019 within the CSM framework~\cite{Lu:2019bjs}--which employed the momentum dependence of the dressed-quark mass deduced from the QCD Lagrangian and successfully described the pion and nucleon elastic form factors, as well as the electrocouplings of the $\Delta(1232)3/2^+$ and $N(1440)1/2^+$--have been confirmed by the CLAS experimental results reported in our previous analysis~\cite{Mokeev:2023zhq} and supported by the current analysis outcome. Therefore, the study of the impact of tails from resonances in the third resonance region on the $\Delta(1600)3/2^+$ electrocouplings support the capability of the CSM approach to provide unique insight into the strong-interaction dynamics that underpin the emergence of the dominant part of hadron mass from the results on the $Q^2$ evolution of the $\gamma_vpN^*$ electrocouplings \cite{Achenbach:2025kfx,Carman:2023zke,Ding:2022ows}.

\subsection{$N(1675)5/2^-$ and $N(1680)5/2^+$ Resonances}
\label{N1675_N1680}

The electroexcitation of the $N(1675)5/2^-$ and $N(1680)5/2^+$ was previously studied in $\pi^+ n$ electroproduction measured with CLAS~\cite{Park:2014yea} within the same $W$ and $Q^2$ ranges covered in the present analysis. For the $N(1675)5/2^-$, the BF for decays into $\pi N$ and $\pi\pi N$ are comparable. In contrast, the $N(1680)5/2^+$ decays preferentially into $\pi N$, with a BF of $60-70$\%, while still exhibiting significant decays into $\pi\pi N$, with a BF of $28-53$\%~\cite{ParticleDataGroup:2024cfk}. The electrocouplings of both resonances, obtained from independent studies of $\pi N$ and $\pi^+\pi^-p$ electroproduction, provide an important consistency check. Moreover, investigations of the $N(1675)5/2^-$ and $N(1680)5/2^+$ in $\pi^+\pi^-p$ electroproduction extend the available information on their total and partial decay widths into $\pi\Delta$ and $\rho p$.

\begin{table*}
\begin{center}
\begin{tabular}{|c|c|c|c|c|c|c|} \hline
                                                & $\Gamma_\text{tot}$ (MeV)  & $\Gamma_{\pi\Delta}$ (MeV) & BF($\pi\Delta$) (\%) & $\Gamma_{\rho p}$ (MeV) & BF($\rho p$) (\%) & Mass (GeV) \\   \hline
$W$: $1.56-1.66$ GeV, $Q^2$: $2.0-3.5$ GeV$^2$  & $147 \pm 6.5$              & $72.9 \pm 7.5$             & $43-57$             & $3.34 \pm 2.58$         & $0.5-4.2$        & $1.675 \pm 0.004$ \\ \hline
$W$: $1.56-1.66$ GeV, $Q^2$: $3.0-5.0$ GeV$^2$  & $146 \pm 8.9$              & $70.6 \pm 8.8$             & $40-58$             & $4.36 \pm 2.97$         & $1.2-5.7$        & $1.673 \pm 0.002$ \\ \hline
$W$: $1.61-1.71$ GeV, $Q^2$: $2.0-3.5$ GeV$^2$  & $147 \pm 9.3$              & $73.5 \pm 8.9$             & $41-60$             & $3.84 \pm 2.80$         & $0.7-4.8$        & $1.673 \pm 0.003$ \\ \hline
$W$: $1.61-1.71$ GeV, $Q^2$: $3.0-5.0$ GeV$^2$  & $144 \pm 9.5$              & $68.1 \pm 10.4$            & $38-58$             & $5.15 \pm 3.60$         & $1.0-6.5$        & $1.677 \pm 0.003$ \\ \hline
$W$: $1.66-1.76$ GeV, $Q^2$: $2.0-3.5$ GeV$^2$  & $146 \pm 8.1$              & $71.3 \pm 9.4$             & $40-59$             & $3.65 \pm 2.99$         & $0.4-4.8$        & $1.674 \pm 0.006$ \\ \hline
$W$: $1.66-1.76$ GeV, $Q^2$: $3.0-5.0$ GeV$^2$  & $148 \pm 9.5$              & $72.9 \pm 8.2$             & $41-59$             & $3.82 \pm 2.53$         & $1.0-6.5$        & $1.677 \pm 0.005$  \\ \hline
 PDG \cite{ParticleDataGroup:2024cfk}           & $130-160$                  &                            & $23-37$             &                         & $0.1-0.9$        & $1.665-1.680$     \\ \hline
\end{tabular}
\caption{The mass, total decay width ($\Gamma_\text{tot}$), partial decay widths ($\Gamma_{\pi\Delta}$, $\Gamma_{\rho p}$), and branching fractions (BF$_{\pi\Delta}$, BF$_{\rho p}$) for the $N(1675)5/2^-$ determined from independent fits of the nine one-fold $\pi^+\pi^-p$ differential cross sections \cite{Trivedi:2026fsd}.}
\label{n1675hadr} 
\end{center}
\end{table*}

\begin{table*}
\begin{center}
\begin{tabular}{|c|c|c|c|c|c|c|} \hline
                                               & $\Gamma_\text{tot}$ (MeV) & $\Gamma_{\pi\Delta}$ (MeV) & BF($\pi\Delta$) (\%) & $\Gamma_{\rho p}$ (MeV) & BF($\rho p$) (\%) & Mass (GeV) \\   \hline
$W$: $1.56-1.66$ GeV, $Q^2$: $2.0-3.5$ GeV$^2$ & $121 \pm 3.2$             & $25.3 \pm 4.1$    & $16-25$  & $19.8 \pm 5.4$  & $12-22$ & $1.685 \pm 0.003$ \\ \hline
$W$: $1.56-1.66$ GeV, $Q^2$: $3.0-5.0$ GeV$^2$ & $122 \pm 4.1$             & $28.4 \pm 4.4$    & $19-28$  & $18.5 \pm 5.8$  & $10-21$ & $1.685 \pm 0.003$ \\ \hline
$W$: $1.61-1.71$ GeV, $Q^2$: $2.0-3.5$ GeV$^2$ & $120 \pm 4.3$             & $24.4 \pm 3.7$    & $17-24$  & $19.6 \pm 6.2$  & $11-22$ & $1.684 \pm 0.003$ \\ \hline
$W$: $1.61-1.71$ GeV, $Q^2$: $3.0-5.0$ GeV$^2$ & $121 \pm 3.4$             & $24.0 \pm 5.4$    & $15-25$  & $21.4 \pm 5.1$  & $13-23$ & $1.686 \pm 0.003$ \\ \hline
$W$: $1.66-1.76$ GeV, $Q^2$: $2.0-3.5$ GeV$^2$ & $122 \pm 4.5$             & $27.6 \pm 5.2$    & $18-28$  & $19.4 \pm 4.4$  & $12-20$ & $1.684 \pm 0.003$ \\ \hline
$W$: $1.66-1.76$ GeV, $Q^2$: $3.0-5.0$ GeV$^2$ & $125 \pm 3.2$             & $31.0 \pm 5.3$    & $20-30$  & $18.5 \pm 6.12$ & $10-20$ & $1.685 \pm 0.003$ \\ \hline
 PDG \cite{ParticleDataGroup:2024cfk}          & $115-130$                 &                   & $11-23$  &                 & $8-11$  & $1.665-1.680$ \\ \hline
\end{tabular}
\caption{The mass, total decay width ($\Gamma_\text{tot}$), partial decay widths ($\Gamma_{\pi\Delta}$, $\Gamma_{\rho p}$), and branching fractions (BF$_{\pi\Delta}$, BF$_{\rho p}$) for the $N(1680)5/2^+$ determined from independent fits of the nine one-fold $\pi^+\pi^-p$ differential cross sections \cite{Trivedi:2026fsd}.}
\label{n1680hadr} 
\end{center}
\end{table*}

The masses, total decay widths ($\Gamma_\text{tot}$), and partial hadronic decay widths $\Gamma_{\pi\Delta}$ and $\Gamma_{\rho p}$ for the $N(1675)5/2^-$ and $N(1680)5/2^+$ were obtained from independent fits to the nine $\pi^+\pi^-p$ one-fold differential cross sections~\cite{Trivedi:2026fsd} within the kinematic regions listed in Table~\ref{Q2vsW_areas}. The extracted parameters are presented in Tables~\ref{n1675hadr} and \ref{n1680hadr}.

The masses, total decay width, and partial decay widths to $\pi\Delta$ and $\rho p$, deduced from independent fits of the $\pi^+\pi^-p$ differential cross sections within the six overlapping $(W,Q^2)$ regions listed in Table~\ref{Q2vsW_areas}, are consistent for both the $N(1675)5/2^-$ and $N(1680)5/2^+$. The extracted parameters agree within uncertainties, indicating a credible extraction from the $\pi^+\pi^-p$ electroproduction data. The masses and total decay widths of both resonances are also consistent with the values reported by the PDG~\cite{ParticleDataGroup:2024cfk}.

The relatively small uncertainties obtained for the masses and total widths of the $N(1675)5/2^-$ and $N(1680)5/2^+$ should, however, be interpreted with caution, since in the fits we varied only the masses and the partial hadronic decay widths to $\pi\Delta$ and $\rho p$, while the contributions from all other decay modes were kept fixed. The branching fractions deduced for the decays of these states into the $\pi\Delta$ and $\rho p$ subchannels are larger than those reported by the PDG. 

Within the broad range $2.0 \leq Q^2 \leq 5.0$~GeV$^2$, the total and partial hadronic decay widths of the $N(1675)5/2^-$ and $N(1680)5/2^+$ remain $Q^2$-independent. This stability provides evidence that both resonances are excited as $s$-channel states in virtual photon-proton interactions, featuring a core of three dressed quarks that can be augmented by an external meson-baryon cloud.

\begin{figure*}[htpb]
\begin{center}
\includegraphics[width=0.9\textwidth]{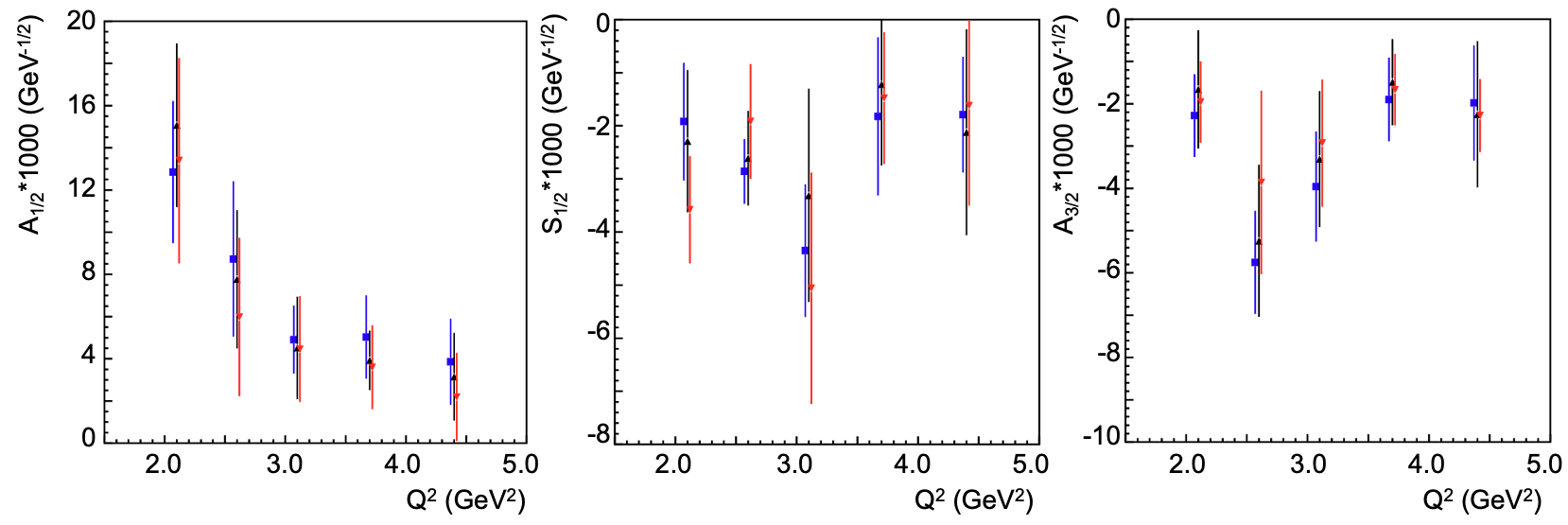}
\caption{Electrocouplings of the $N(1675)5/2^-$, $A_{1/2}$ (left), $S_{1/2}$ (center), and $A_{3/2}$ (right), determined in the independent analyses across three $W$ intervals $1.56-1.66$~GeV (blue), $1.61-1.71$~GeV (black), and $1.66-1.76$~GeV (red).}  
\label{3W_n1675}
\end{center}
\end{figure*}

\begin{figure*}[htbp]
\begin{center}
\includegraphics[width=0.9\textwidth]{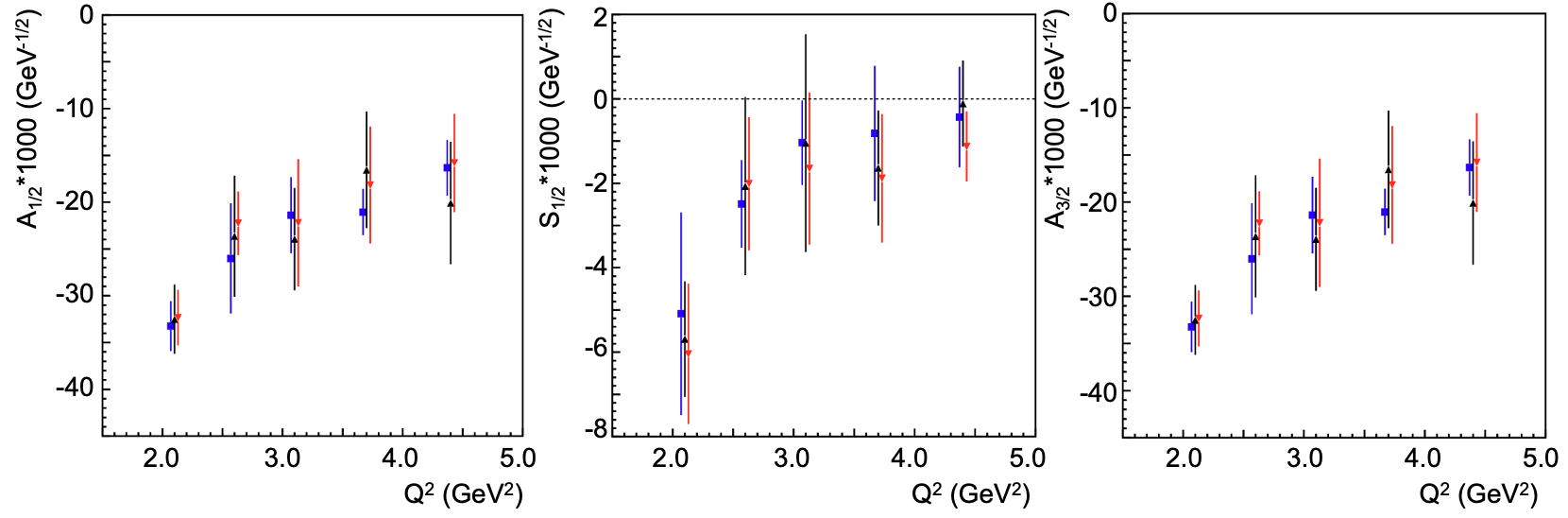}
\caption{Electrocouplings of the $N(1680)5/2^+$, $A_{1/2}$ (left), $S_{1/2}$ (center), and $A_{3/2}$ (right), determined in the independent analyses across three $W$ intervals $1.56-1.66$~GeV (blue), $1.61-1.71$~GeV (black), and $1.66-1.76$~GeV (red).}  
\label{3W_n1680}
\end{center}
\end{figure*}

The electrocouplings of the $N(1675)5/2^-$ and $N(1680)5/2^+$ were extracted from independent fits to the nine one-fold $\pi^+\pi^-p$ differential cross sections across the three overlapping $W$ intervals listed in Table~\ref{Q2vsW_areas}. The results, presented in Figs.~\ref{3W_n1675} and \ref{3W_n1680}, are seen to be consistent across the three $W$ intervals. In fact, within uncertainties, the results are nearly coincident across the entire $Q^2$ range covered by the measurements~\cite{Trivedi:2026fsd}. Although the non-resonant mechanisms of the JM23 reaction model differ across the three $W$ intervals, the consistent extraction of the electrocouplings in all cases provides strong evidence for the reliable isolation of the resonant amplitudes, and thus the credible determination of the $N(1675)5/2^-$ and $N(1680)5/2^+$ electrocouplings from $\pi^+\pi^-p$ electroproduction data within the JM23 framework.

\begin{figure*}[htp]
\begin{center}
\includegraphics[width=0.9\textwidth]{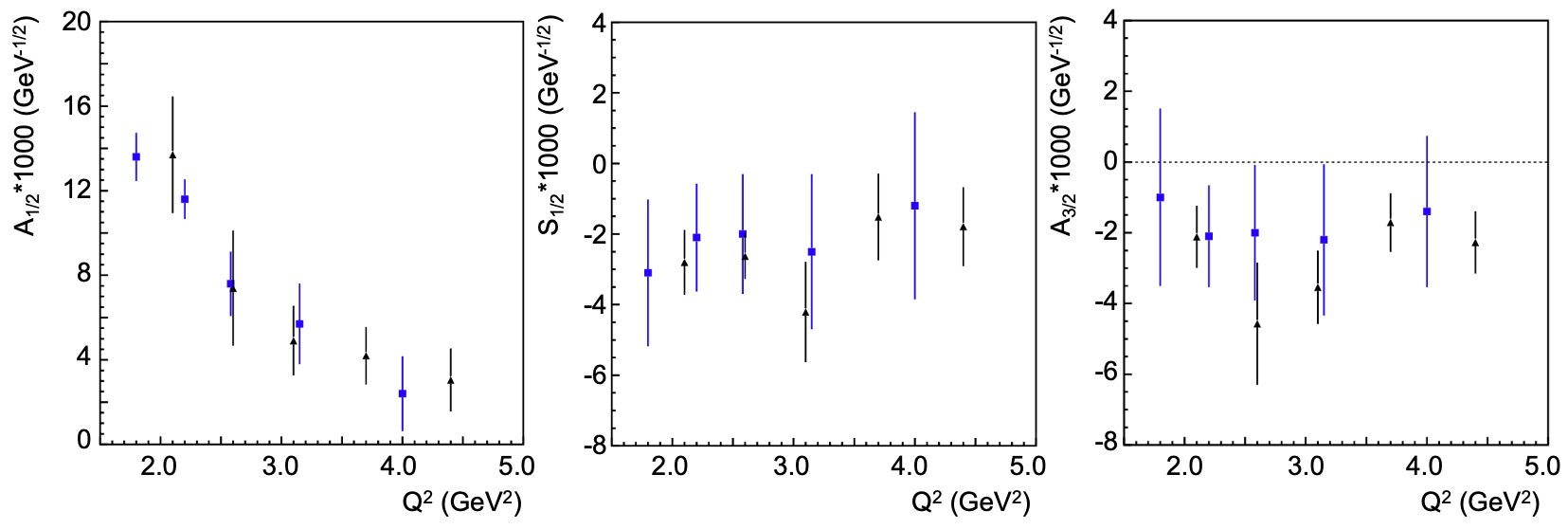}
\caption{Electrocouplings of the $N(1675)5/2^-$, $A_{1/2}$ (left), $S_{1/2}$ (center), and $A_{3/2}$ (right), determined in the analyses of $\pi^+\pi^-p$~\cite{Trivedi:2026fsd} (black) and $\pi N$~\cite{Park:2014yea} CLAS data (blue). The results from the $\pi^+\pi^-p$ channel are averaged across three $W$ intervals $1.56-1.66$~GeV, $1.61-1.71$~GeV, and $1.66-1.76$~GeV.}  
\label{n1675el}
\end{center}
\end{figure*}

\begin{figure*}[htpb]
\begin{center}
\includegraphics[width=0.9\textwidth]{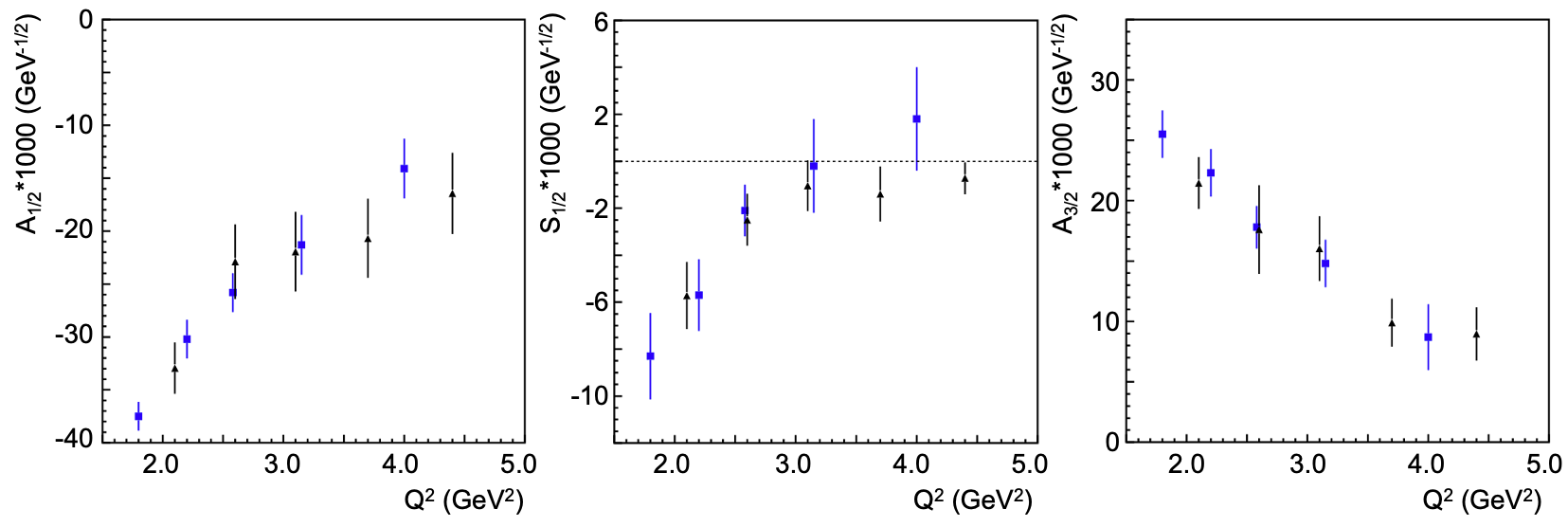}
\caption{Electrocouplings of the $N(1680)5/2^+$, $A_{1/2}$ (left), $S_{1/2}$ (center), and $A_{3/2}$ (right), determined in the analyses of $\pi^+\pi^-p$~\cite{Trivedi:2026fsd} (black) and $\pi N$~\cite{Park:2014yea} CLAS data (blue). The results from the $\pi^+\pi^-p$ channel are averaged across three $W$ intervals $1.56-1.66$~GeV, $1.61-1.71$~GeV, and $1.66-1.76$~GeV.}  
\label{n1680el}
\end{center}
\end{figure*}

The averages of the $\gamma_vpN^*$ electrocouplings for the $N(1675)5/2^-$ and $N(1680)5/2^+$ obtained across the three $W$ intervals are shown in Figs.~\ref{n1675el} and \ref{n1680el} as a function of $Q^2$. These results are compared with the electrocouplings deduced from $\pi N$ electroproduction data measured with CLAS~\cite{Park:2014yea}. The averaging procedure is described in the introductory part of Section~\ref{elcoupl_hadrdec}. The numerical results on the electrocouplings averaged over the three $W$-intervals for the $N(1675)5/2^-$ and $N(1680)5/2^+$ are shown in Tables~\ref{n1675el_final} and \ref{n1680el_final}.

\begin{table}
\begin{center}
\begin{tabular}{|c|c|c|c|} \hline
$Q^2$ Interval  & $A_{1/2} \times 1000$ & $S_{1/2} \times 1000$ & $A_{3/2} \times 1000$  \\
 (GeV$^2$)      & (GeV$^{-1/2}$)        & (GeV$^{-1/2}$)        &  (GeV$^{-1/2}$)        \\ \hline
 $2.0-2.4$      &  $13.70 \pm 2.76$     & $-2.80 \pm 0.92$ &  $-2.11 \pm 0.88$ \\ \hline
 $2.4-3.0$      &  $7.39 \pm 2.73$      & $-2.63 \pm 0.64$ &  $-4.57 \pm 1.73$ \\ \hline
 $3.0-3.5$      &  $4.91 \pm 1.65$      & $-4.21 \pm 1.42$ &  $-3.54 \pm 1.04$ \\ \hline
 $3.5-4.2$      &  $4.20 \pm 1.36$      & $-1.52 \pm 1.23$ &  $-1.71 \pm 0.83$ \\ \hline
 $4.2-5.0$      &  $3.05 \pm 1.4$9      & $-1.79 \pm 1.12$ &  $-2.27 \pm 0.88$ \\ \hline
\end{tabular}
\caption{$N(1675)5/2^-$ electrocouplings determined from the fits of $\pi^+\pi^-p$ differential cross sections \cite{Trivedi:2026fsd} and averaged across three $W$ intervals, $1.56-1.66$~GeV, $1.61-1.71$~GeV, and $1.66-1.76$~GeV, for $Q^2$ from $2.0-5.0$~GeV$^2$.} 
\label{n1675el_final}
\end{center}
\end{table}

\begin{table}
\begin{center}
\begin{tabular}{|c|c|c|c|}
\hline
$Q^2$ Interval  & $A_{1/2} \times 1000$ & $S_{1/2} \times 1000$ & $A_{3/2} \times 1000$  \\
 (GeV$^2$)      & (GeV$^{-1/2}$)        & (GeV$^{-1/2}$)        &  (GeV$^{-1/2}$)        \\ \hline
 $2.0-2.4$        &  $-32.94 \pm 2.42$    & $-5.72 \pm 1.43$      &  $21.46 \pm 2.15$ \\ \hline
 $2.4-3.0$        &  $-22.88 \pm 3.53$    & $-2.49 \pm 1.11$      &  $17.61 \pm 3.67$ \\ \hline
 $3.0-3.5$        &  $-21.94 \pm 3.76$    & $-1.04 \pm 1.08$      &  $16.04 \pm 2.68$ \\ \hline
 $3.5-4.2$        &  $-20.68 \pm 3.74$    & $-1.39 \pm 1.17$      &  $9.90 \pm 1.99$ \\ \hline
 $4.2-5.0$        &  $-16.43 \pm 3.84$    & $-0.72 \pm 0.68$      &  $8.97 \pm 2.21$ \\ \hline
\end{tabular}
\caption{$N(1680)5/2^+$ electrocouplings determined from the fits of $\pi^+\pi^-p$ differential cross sections \cite{Trivedi:2026fsd} and averaged across three $W$ intervals, $1.56-1.66$~GeV, $1.61-1.71$~GeV, and $1.66-1.76$~GeV, for $Q^2$ from $2.0-5.0$~GeV$^2$.} 
\label{n1680el_final}
\end{center}
\end{table}

The results on the electrocouplings of the $N(1675)5/2^-$ and $N(1680)5/2^+$ obtained from independent studies of the $\pi N$ and $\pi^+\pi^-p$ electroproduction channels are consistent across the broad range of photon virtualities, $2.0 \leq Q^2 \leq 5.0$~GeV$^2$, covered by the CLAS measurements. While the non-resonant mechanisms in the $\pi N$ and $\pi^+\pi^-p$ channels are different, the consistent results for both resonances extracted from these two exclusive channels--the dominant contributors in the resonance region--provide conclusive evidence for the capabilities of the reaction models developed for $\pi N$ \cite{Aznauryan:2002gd,Aznauryan:2009mx,Park:2014yea} and $\pi^+\pi^-p$~\cite{Mokeev:2008iw, Mokeev:2012vsa, Mokeev:2015lda} electroproduction to credibly extract these quantities.

The $N(1675)5/2^-$ excited off protons exhibits peculiar structural features. In the limit of exact $SU(6)\otimes O(3)$ symmetry for the resonance wavefunction and under the dominance of a single-quark transition current, the transverse amplitudes $A_{1/2}$ and $A_{3/2}$ are expected to vanish~\cite{Hey:1974qe, Babcock:1975bw, Cottingham:1978za, Burkert:2002zz}. Breaking of $SU(6)\otimes O(3)$ symmetry through mixing of three-quark configurations generates contributions from the quark core~\cite{Aznauryan:1985, Santopinto:2012nq, Merten:2003iy}. However, the quark-core contributions predicted in various quark models remain far smaller than the electrocouplings of the $N(1675)5/2^-$ extracted in previous $\pi^+n$ electroproduction studies~\cite{Aznauryan:2014xea}.

The meson-baryon cloud contributions to the electrocouplings for the $N(1675)5/2^-$, evaluated within the Argonne-Osaka coupled-channel analysis at the pole position~\cite{Julia-Diaz:2007mae}, are comparable in magnitude to the absolute values of the $A_{1/2}$ and $A_{3/2}$ amplitudes measured in CLAS electroexcitation studies over the range $2.0 \leq Q^2 \leq 5.0$~GeV$^2$. This demonstrates that meson-baryon cloud contributions dominate the structure of the $N(1675)5/2^-$ even at the highest $Q^2$ values probed.

The consistent results on the $N(1675)5/2^-$ electrocouplings obtained independently from the $\pi^+n$ and $\pi^+\pi^-p$ electroproduction channels reported here for the first time provide a promising opportunity to deepen our understanding of the meson-baryon cloud. These studies may yield valuable insight into the dynamics of the transition between the confined dressed quark core and the meson-baryon degrees of freedom associated with the cloud, thereby addressing a key open problem related to the nature of confinement.

\subsection{$\Delta(1700)3/2^-$ Resonance}
\label{Delta1700)}

According to the PDG~\cite{ParticleDataGroup:2024cfk}, the $\Delta(1700)3/2^-$ decays predominantly into $\pi\Delta$ and $\rho p$. This makes the $\pi^+\pi^-p$ electroproduction data the primary source of information on the parameters of this state. The PDG reports its Breit-Wigner mass in the range $1.69-1.73$~GeV. Accounting for the decay width of this state, we have determined its parameters by fitting the nine one-fold differential cross sections across two overlapping $W$ intervals, $1.61-1.71$~GeV and $1.66-1.76$~GeV, for two overlapping $Q^2$ intervals, $2.0-3.5$~GeV$^2$ and $3.0-5.0$~GeV$^2$. The resonance mass, the total decay width ($\Gamma_\text{tot}$), and the partial decay widths to the $\pi\Delta$ ($\Gamma_{\pi\Delta}$) and $\rho p$ ($\Gamma_{\rho p}$) final states obtained from these fits are shown in Table~\ref{delta1700hadr}.

\begin{table*}
\begin{center}
\begin{tabular}{|c|c|c|c|c|c|c|} \hline
                                            & $\Gamma_\text{tot}$ (MeV) & $\Gamma_{\pi\Delta}$ (MeV) & BF($\pi\Delta$) (\%) & $\Gamma_{\rho p}$ (MeV) & BF($\rho p$) (\%) & Mass (GeV) \\   \hline
$W$: $1.61-1.71$ GeV, $Q^2$: $2.0-3.5$ GeV$^2$  & $304 \pm 22$        & $290 \pm 16$              & $>83$              & $14.3 \pm 7.3$         & $2.1-7.7$          & $1.707 \pm 0.008$ \\ \hline     
$W$: $1.61-1.71$ GeV, $Q^2$: $3.0-5.0$ GeV$^2$  & $303 \pm 36$        & $287 \pm 36$              & $>74$              & $15.4 \pm 4.6$         & $3.2-7.5$          & $1.716 \pm 0.011$ \\ \hline
$W$: $1.66-1.76$ GeV, $Q^2$: $2.0-3.5$ GeV$^2$  & $295 \pm 20$        & $276 \pm 19$              & $>82$              & $16.2 \pm 7.3$         & $2.8-8.5$          & $1.708 \pm 0.013$ \\ \hline
$W$: $1.66-1.76$ GeV, $Q^2$: $3.0-5.0$ GeV$^2$  & $310 \pm 27$        & $299 \pm 27$              & $>81$              & $11.5 \pm 1.1$         & $3.1-4.4$          & $1.708 \pm 0.015$ \\ \hline
PDG \cite{ParticleDataGroup:2024cfk}            & $220-380$           &                           & $9-70$             &                        & $22-32$            & $1.690-1.730$ \\ \hline
\end{tabular}
\caption{The mass, total decay width ($\Gamma_\text{tot}$), partial decay widths ($\Gamma_{\pi\Delta}$, $\Gamma_{\rho p}$), and branching fractions (BF$_{\pi\Delta}$, BF$_{\rho p}$) for the $\Delta(1700)3/2^-$ determined from independent fits of the nine one-fold $\pi^+\pi^-p$ differential cross sections \cite{Trivedi:2026fsd}.}
\label{delta1700hadr} 
\end{center}
\end{table*}

The parameters determined from the independent fits within the overlapping $(W,Q^2)$ intervals are consistent within their uncertainties. No evidence is observed for any evolution of these parameters with $Q^2$, suggesting that the $\Delta(1700)3/2^-$ is an $s$-channel resonance excited in virtual photon-proton interactions. This state must therefore possess an inner core of three dressed quarks, which may be augmented by an external meson-baryon cloud. The mass and total decay width of the $\Delta(1700)3/2^-$ inferred from our $\pi^+\pi^-p$ fits lie within the relatively large uncertainties reported by the PDG \cite{ParticleDataGroup:2024cfk}. However, the branching fractions for its hadronic decays into $\pi\Delta$ and $\rho p$ deduced in our analysis are larger than those listed by the PDG.

\begin{figure*}
\begin{center}
\includegraphics[width=0.9\textwidth]{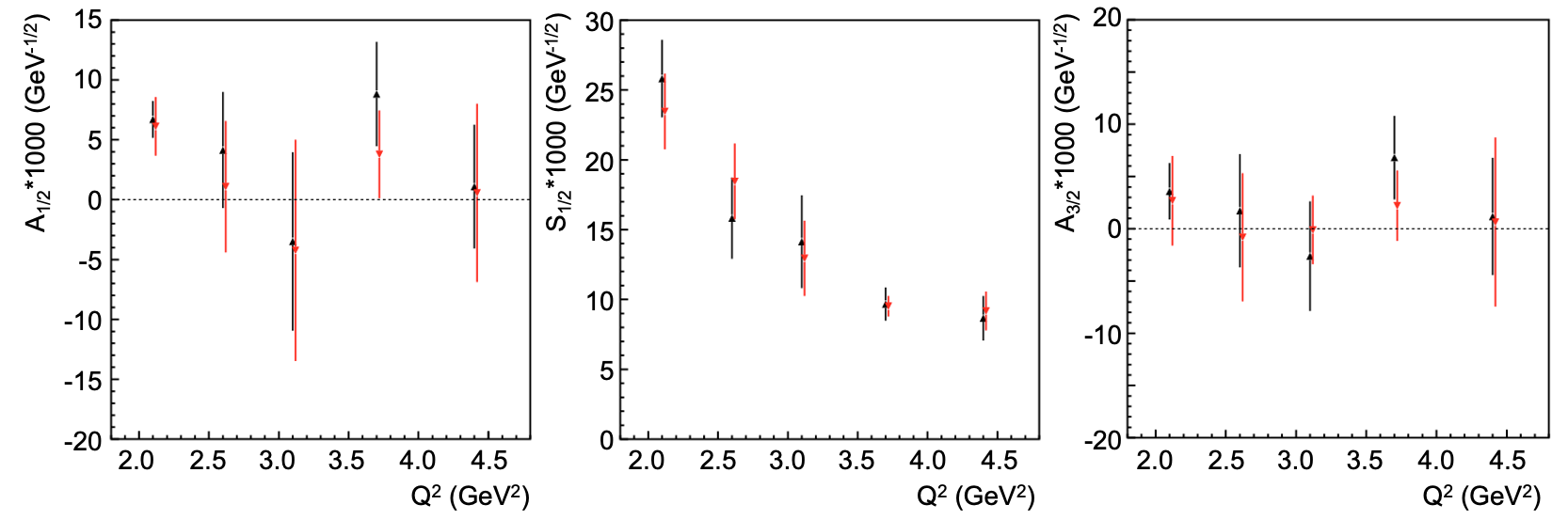}
\caption{Electrocouplings of the $\Delta(1700)3/2^-$, $A_{1/2}$ (left), $S_{1/2}$ (center), and $A_{3/2}$ (right), determined in independent fits of $\pi^+\pi^-p$ CLAS data~\cite{Trivedi:2026fsd} across two overlapping $W$-intervals $1.61-1.71$~GeV (black) and $1.66-1.76$~GeV (red) for $Q^2$ from $2.0-5.0~$GeV$^2$.}  
\label{delta1700el}
\end{center}
\end{figure*}

The electrocouplings of the $\Delta(1700)3/2^-$, obtained from our $\pi^+\pi^-p$ fits across the two overlapping $W$ intervals ($1.61-1.71$~GeV and $1.66-1.76$~GeV) for $2.0 < Q^2 < 5.0$~GeV$^2$, are shown in Fig.~\ref{delta1700el}. These results on the $Q^2$ evolution of the $\Delta(1700)3/2^-$ electrocouplings are consistent over the entire $Q^2$ range covered. The non-resonant amplitudes in the two $W$ intervals are different and exhibit differences in their $Q^2$ dependence. These results highlight the capability of the JM23 model to reliably extract the electrocouplings of excited nucleon states in the third resonance region from $\pi^+\pi^-p$ differential cross section data.

\begin{figure*}
\begin{center}
\includegraphics[width=0.9\textwidth]{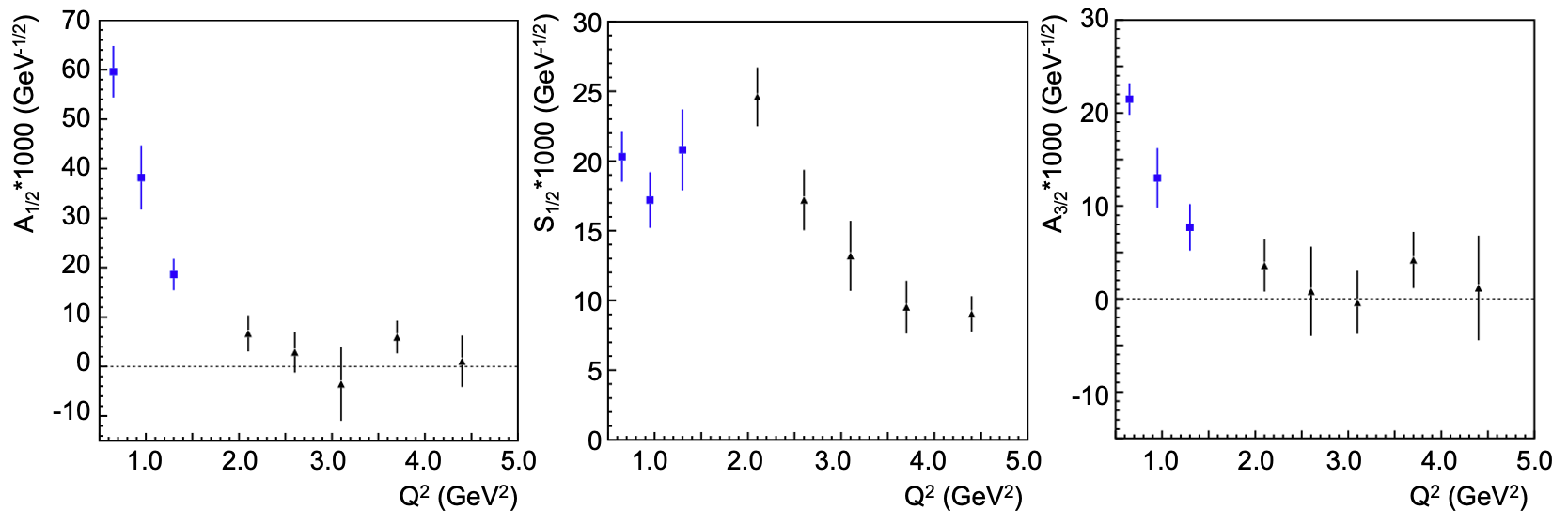}
\caption{Electrocouplings of the $\Delta(1700)3/2^-$, $A_{1/2}$ (left), $S_{1/2}$ (center), and $A_{3/2}$ (right), determined in the analyses of $\pi^+\pi^-p$ differential cross sections~\cite{Trivedi:2026fsd} (black) and the previous results for $Q^2 < 2.0$~GeV$^2$ \cite{Mokeev:2020hhu} (blue). The results for $Q^2 > 2.0$~GeV$^2$ are averaged across the two $W$ intervals $1.61-1.71$~GeV and $1.66-1.76$~GeV.}  
\label{delta1700elav}
\end{center}
\end{figure*}

The results of the independent fits across the two overlapping $W$ intervals were averaged as described in Section~\ref{elcoupl_hadrdec}. The $Q^2$ evolution of the averaged values of the $\Delta(1700)3/2^-$ electrocouplings is shown in Fig.~\ref{delta1700elav}, together with previously published results for this state obtained from $\pi^+\pi^-p$ electroproduction off protons at $Q^2 < 2.0$~GeV$^2$ \cite{Mokeev:2020hhu,CLAS:2002xbv}.

The electroexcitation of the $\Delta(1700)3/2^-$ for $Q^2$ $>$ 2.0~GeV$^2$ is dominated by the longitudinal $S_{1/2}$ electrocouplings, while the transverse $A_{1/2}$ and $A_{3/2}$ electrocouplings are comparable with zero. The numerical results on the $\Delta(1700)3/2^-$ electrocouplings averaged over the two $W$-intervals are shown in Table~\ref{delta1700el_final}.

\begin{table}
\begin{center}
\begin{tabular}{|c|c|c|c|}
\hline
$Q^2$ Interval & $A_{1/2} \times 1000$ & $S_{1/2} \times 1000$ & $A_{3/2} \times 1000$  \\
 (GeV$^2$)     & (GeV$^{-1/2}$)        & (GeV$^{-1/2}$)        & (GeV$^{-1/2}$)  \\ \hline
 $2.0-2.4$     & $6.70 \pm 3.07$       & $24.6 \pm 2.1$        & $3.58 \pm 2.81$ \\ \hline
 $2.4-3.0$     & $2.92 \pm 4.13$       & $17.2 \pm 2.2$        & $0.81 \pm 4.80$ \\ \hline
 $3.0-3.5$     & $-3.49 \pm 7.48$      & $13.2 \pm 2.5$        & $-0.38 \pm 3.39$ \\ \hline
 $3.5-4.2$     & $5.97 \pm 3.31$       & $9.52 \pm 1.89$       & $4.18 \pm 3.03$ \\ \hline
 $4.2-5.0$     & $1.08 \pm 5.19$       & $9.03 \pm 1.27$       & $1.17 \pm 5.63$ \\ \hline
\end{tabular}
\caption{$\Delta(1700)3/2^-$ electrocouplings determined from fits of the $\pi^+\pi^-p$ differential cross sections \cite{Trivedi:2026fsd} and averaged across the two $W$ intervals $1.61-1.71$~GeV and $1.66-1.76$~GeV for $Q^2$ from $2.0-5.0$~GeV$^2$.} 
\label{delta1700el_final}
\end{center}
\end{table}

\subsection{Insight into DCSB in Connection with EHM}
\label{DCSB_EHM}

The results on the $\gamma_v p N^*$ electrocouplings of the chiral-partner states $N(1675)5/2^-$ and $N(1680)5/2^+$, together with those for the $\Delta(1700)3/2^-$ reported here and the previously studied electrocouplings of the $\Delta(1232)3/2^+$ in $\pi N$ electroproduction~\cite{Aznauryan:2009mx}, open new opportunities to explore the impact of dynamical chiral symmetry breaking (DCSB) on the structure of chiral-partner resonances. Within the framework of CSM, the concept of EHM has revealed a deep connection between two fundamental features of strongly coupled QCD-DCSB and EHM \cite{Achenbach:2025kfx,Ding:2022ows,Carman:2023zke,Horn:2016rip, Cheng:2025sdp}. This connection can now be further tested through the experimental results on the electrocouplings of chiral-partner resonance pairs.

According to the EHM concept developed in the chiral limit of massless bare quarks, the dynamically generated masses of dressed quarks $q$ and $\bar{q}$ coupled to pseudoscalar mesons are exactly canceled by the $q\bar{q}$ interaction, resulting in a massless octet of pseudoscalar mesons with identical internal structure~\cite{Achenbach:2025kfx,Ding:2022ows,Horn:2016rip, Roberts:2016vyn}. When chiral symmetry is broken explicitly by the nonzero current-quark masses generated through the Higgs mechanism, and implicitly by the emergent contributions to the dressed-quark masses, the $q\bar{q}$ interaction reduces the dynamically generated quark masses in bound $q\bar{q}$ systems with $J^P=0^-$, but not all the way to zero. Consequently, pseudoscalar mesons acquire nonzero masses that remain smaller than those of other non-strange and strange mesons. In this way, CSM explains the dual nature of the pseudoscalar octet: they are simultaneously bound $q\bar{q}$ systems and the Goldstone bosons of DCSB. In the meson sector, DCSB manifests itself in the structural features of pseudoscalar color-singlet mesons. In baryons, including the $N^*$ states, the impact of DCSB manifests itself in the generation of correlated $qq$ color antitriplet diquark systems of opposite parity to the $q\bar{q}$ meson states~\cite{Cahill:1987qr, Barabanov:2020jvn, Cheng:2025yij}, as well as in the generation of the emergent part of the dressed quark mass.

The successful description of the pion and nucleon elastic electromagnetic form factors, together with the electrocouplings of resonances of different structure--$\Delta(1232)3/2^+$, $N(1440)1/2^+$, and $\Delta(1600)3/2^+$--using the same momentum dependence of the dressed-quark mass~\cite{Achenbach:2025kfx}, provides strong evidence for insight into EHM. To shed light on the connection between EHM and DCSB, comparative analyses of our results on the $Q^2$ evolution of the electrocouplings for pairs of chiral-partner resonances are needed.

The Breit-Wigner masses of the chiral partners $N(1675)5/2^-$ and $N(1680)5/2^+$ are nearly identical; however, this alone does not provide evidence for proximity to chiral symmetry restoration. The results on the $Q^2$ evolution of their electrocouplings, obtained from studies of $\pi N$~\cite{Park:2014yea} and $\pi^+\pi^-p$ (this work) electroproduction and shown in Figs.~\ref{n1675el} and \ref{n1680el}, reveal pronounced differences. For the $N(1675)5/2^-$, the transverse $A_{1/2}$ amplitude is the dominant contributor at $Q^2 < 3.5$~GeV$^2$, while at higher $Q^2$ all three electrocouplings approach nearly equal absolute values. In contrast, for the chiral partner $N(1680)5/2^+$, the $A_{1/2}$ amplitude remains the biggest contribution across the entire $Q^2$ range covered in the CLAS measurements. In both cases, the absolute magnitudes of the transverse electroexcitation amplitudes substantially exceed those of the longitudinal amplitudes.

The results on the $Q^2$ evolution of the $\Delta(1700)3/2^-$ electrocouplings, deduced in the present analysis of $\pi^+\pi^-p$ electroproduction data and shown in Fig.~\ref{delta1700elav}, exhibit marked differences compared with previous CLAS results on the electrocouplings of its chiral partner $\Delta(1232)3/2^+$, displayed in Fig.~9 (top) of Ref.~\cite{Achenbach:2025kfx}. The electroexcitation of the $\Delta(1232)3/2^+$ is dominated by the transverse $A_{1/2}$ and $A_{3/2}$ amplitudes, with the largest contribution arising from $A_{3/2}$. In contrast, the electroexcitation of the chiral partner $\Delta(1700)3/2^-$ is governed by the longitudinal $S_{1/2}$ amplitude across the entire $Q^2$ range from 2.0 to 5.0~GeV$^2$. The dominance of $S_{1/2}$ represents a striking and peculiar feature for $N^*$ electroexcitation.

The pronounced differences observed in the $Q^2$ evolution of the electrocouplings of the chiral partner pairs $N(1675)5/2^-$-$N(1680)5/2^+$ and $\Delta(1232)3/2^+$-$\Delta(1700)3/2^-$ offer a promising opportunity to further understand the manifestation of DCSB in connection with EHM. In particular, CSM studies can extend insight into the underlying $qq$-correlation dynamics, enabling exploration of the extent to which correlations of opposite parity contribute to the structural differences between chiral partner resonances, as anticipated in earlier CSM work~\cite{Lu:2017cln}. The electroexcitation of the $\Delta(1232)3/2^+$ has been extensively studied within the CSM~\cite{Segovia:2014aza}, while the exploratory CSM evaluation of the electrocouplings of its chiral partner $\Delta(1700)3/2^-$ employing a simplified contact $qq$-interaction resulting in a momentum-independent dressed quark mass has become available~\cite{Albino:2025fcp}. The predicted $\Delta(1700)3/2^-$ $S_{1/2}$ electrocoupling values are far from our results, which indicates an important role of the running dressed quark mass. The $\Delta(1700)3/2^-$ electrocouplings have recently been evaluated with a realistic, QCD-connected $qq$-interaction that generates a running dressed quark mass and $qq$-correlation amplitudes that account for the connection between EHM and DCSB~\cite{Cheng:2025sdp}. A good description of our results has been achieved for $Q^2 > 2.0$~GeV$^2$, and thus, a connection between DCSB and EHM has been observed for the first time in baryon sector.

\subsection{$N(1720)3/2^+$ and $N'(1720)3/2^+$ Resonances}
\label{sec:N1720}

The new $N'(1720)3/2^+$ state was observed in combined analyses of $\pi^+\pi^-p$ photo- and electroproduction data for $Q^2 < 1.5$~GeV$^2$~\cite{Mokeev:2020hhu}. Contributions from this state, together with those from the well-established $N(1720)3/2^+$ state, were required to achieve a successful description of the $\pi^+\pi^-p$ differential cross sections in both photo- and electroproduction. This description was obtained with $Q^2$-independent values of the mass, total decay width, and partial decay widths to $\pi \Delta$ and $\rho p$ for both states, as well as for other states contributing to the resonance peak in the third resonance region, clearly visible in the $W$-dependence of the fully integrated $\pi^+\pi^-p$ electroproduction cross sections.

Although the masses of the $N(1720)3/2^+$ and $N'(1720)3/2^+$ are close, they are not identical. More importantly, their hadronic decay widths into $\pi \Delta$ and $\rho p$, along with the $Q^2$-evolution of their electrocouplings, are distinctly different. These differences provided the evidence necessary to establish the existence of both states.

The analysis of the CLAS $\pi^+\pi^-p$ electroproduction data in the range $2.0 < Q^2 < 5.0$~GeV$^2$~\cite{Trivedi:2026fsd}, carried out within the JM23 model~\cite{Mokeev:2023zhq} and presented here, demonstrates that both the new $N'(1720)3/2^+$ and the conventional $N(1720)3/2^+$ contribute to the $\pi^+\pi^-p$ differential cross sections.

The Breit-Wigner mass, total decay width ($\Gamma_\text{tot}$), and partial hadronic decay widths to $\pi \Delta$ ($\Gamma_{\pi \Delta}$) and $\rho p$ ($\Gamma_{\rho p}$) for the $N'(1720)3/2^+$, as determined from the fit to the $\pi^+\pi^-p$ differential cross sections, are summarized in Table~\ref{Np1720hadr}. The masses and total/partial decay widths of the $N'(1720)3/2^+$ obtained from independent analyses of the two overlapping $W$ intervals, $1.61-1.71$~GeV and $1.66-1.76$~GeV, are consistent with each other. They are also in agreement with the values determined in previous studies of the $N'(1720)3/2^+$ for $0.0 < Q^2 < 1.5$~GeV$^2$~\cite{Mokeev:2020hhu}.

\begin{table*}
\begin{center}
\begin{tabular}{|c|c|c|c|c|c|c|} \hline
                                            & $\Gamma_\text{tot}$ (MeV) & $\Gamma_{\pi\Delta}$ (MeV) & BF($\pi\Delta$) (\%) & $\Gamma_{\rho p}$ (MeV) & BF($\rho p$) (\%) & Mass (GeV) \\   \hline
$W$: $1.61-1.71$ GeV, $Q^2$: $2.0-3.5$ GeV$^2$  & $119.9 \pm 3.3$     & $69.4 \pm 3.7$            & $56-63$               & $4.95 \pm 0.93$        & $3.2-5.0$          & $1.727 \pm 0.008$ \\ \hline     
$W$: $1.61-1.71$ GeV, $Q^2$: $3.0-5.0$ GeV$^2$  & $119.5 \pm 2.9$     & $67.8 \pm 2.6$            & $53-68$               & $6.3 \pm 1.7$          & $3.8-6.8$          & $1.725 \pm 0.008$ \\ \hline
$W$: $1.66-1.76$ GeV, $Q^2$: $2.0-3.5$ GeV$^2$  & $119.5 \pm 4.5$     & $68.4 \pm 4.1$            & $53-63$               & $5.6 \pm 1.7$          & $3.2-5.0$          & $1.722 \pm 0.006$ \\ \hline
$W$: $1.66-1.76$ GeV, $Q^2$: $3.0-5.0$ GeV$^2$  & $118.1 \pm 3.2$     & $65.9 \pm 2.9$            & $52-59$               & $6.7 \pm 2.4$          & $3.5-7.9$          & $1.723 \pm 0.005$ \\ \hline
$W$: $1.61-1.76$ GeV, $Q^2$: $0.0-1.5$ GeV$^2$  & $119 \pm 7$         &                           & $47-64$               &                        & $3-10$             & $1.725 \pm 0.010$ \\ \hline
\end{tabular}
\caption{The mass, total decay width ($\Gamma_\text{tot}$), partial decay widths ($\Gamma_{\pi\Delta}$, $\Gamma_{\rho p}$), and branching fractions (BF$_{\pi\Delta}$, BF$_{\rho p}$) for the $N'(1720)3/2^+$ determined from independent fits of the nine one-fold $\pi^+\pi^-p$ differential cross sections \cite{Trivedi:2026fsd}. The results from our previous publication~\cite{Mokeev:2020hhu} with the $N'(1720)3/2^+$ parameters determined from the $\pi^+\pi^-p$ photo- and electroproduction differential cross sections for $Q^2$ up to 1.5~GeV$^2$ are presented for comparison in the bottom row.}
\label{Np1720hadr} 
\end{center}
\end{table*}

\begin{table}
\begin{center}
\begin{tabular}{|c|c|c|c|}
\hline
$Q^2$ Interval & $A_{1/2} \times 1000$ & $S_{1/2} \times 1000$ & $A_{3/2} \times 1000$  \\
 (GeV$^2$)     & (GeV$^{-1/2}$)        & (GeV$^{-1/2}$)        & (GeV$^{-1/2}$) \\ \hline
 $2.0-2.4$     & $45.8 \pm 9.1$        & $13.4 \pm 2.5$        & $-13.0 \pm 2.8$ \\ \hline
 $2.4-3.0$     & $44.9 \pm 5.5$        & $14.2 \pm 2.5$        & $-9.1 \pm 2.8$ \\ \hline
 $3.0-3.5$     & $42.2 \pm 8.7$        & $7.8 \pm 2.4$         & $-5.7 \pm 2.1$ \\ \hline
 $3.5-4.2$     & $26.5 \pm 8.4$        & $9.9 \pm 2.2$         & $-9.2 \pm 3.9$ \\ \hline
 $4.2-5.0$     & $18.4 \pm 3.8$        & $2.7 \pm 1.5$         & $-5.7 \pm 4.4$  \\ \hline
\end{tabular}
\caption{$N'(1720)3/2^+$ electrocouplings determined from fits of the $\pi^+\pi^-p$ differential cross sections \cite{Trivedi:2026fsd} averaged over the $W$ intervals $1.61-1.71$~GeV and $1.66-1.76$~GeV for $Q^2$ from $2.0-5.0$~GeV$^2$.} 
\label{Np1720el_final}
\end{center}
\end{table}

\begin{figure*}
\begin{center}
\includegraphics[width=0.9\textwidth]{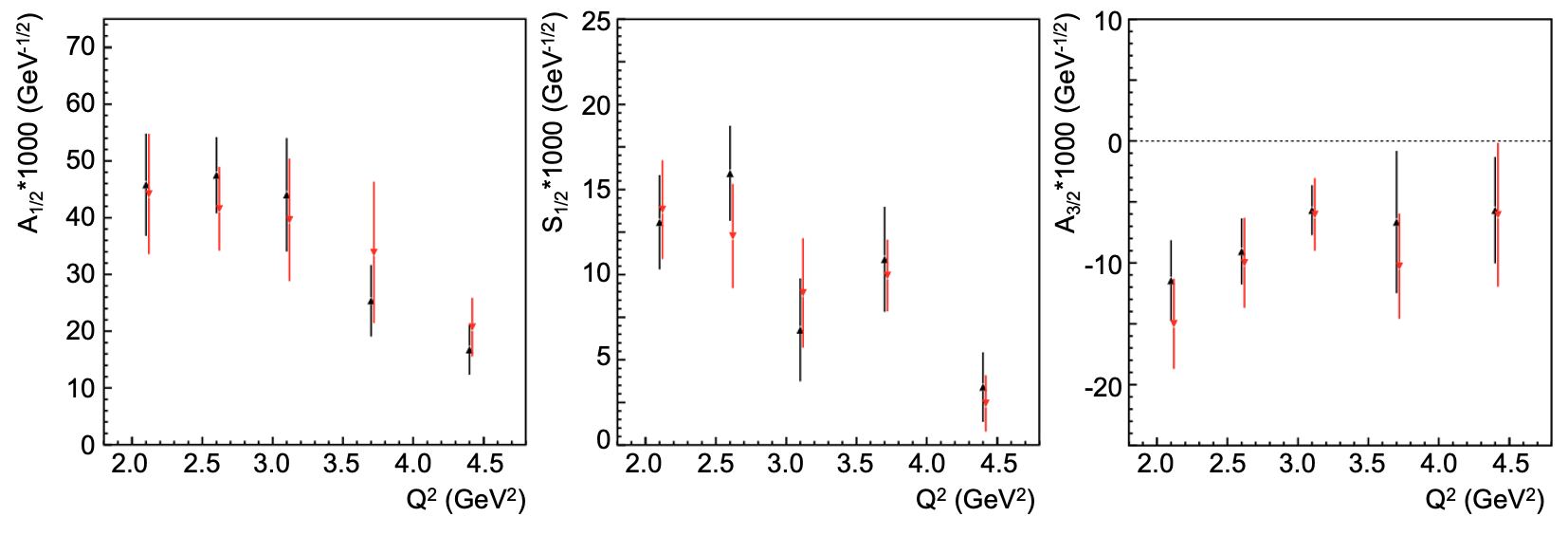}
\caption{Electrocouplings of the $N'(1720)3/2^+$, $A_{1/2}$ (left), $S_{1/2}$ (center), and $A_{3/2}$ (right), determined in independent fits of $\pi^+\pi^-p$ CLAS data~\cite{Trivedi:2026fsd} across two overlapping $W$-intervals $1.61-1.71$~GeV (black) and $1.66-1.76$~GeV (red) for $Q^2$ from $2.0-5.0$~GeV$^2$.}  
\label{Np1720el}
\end{center}
\end{figure*}

\begin{figure*}
\begin{center}
\includegraphics[width=0.9\textwidth]{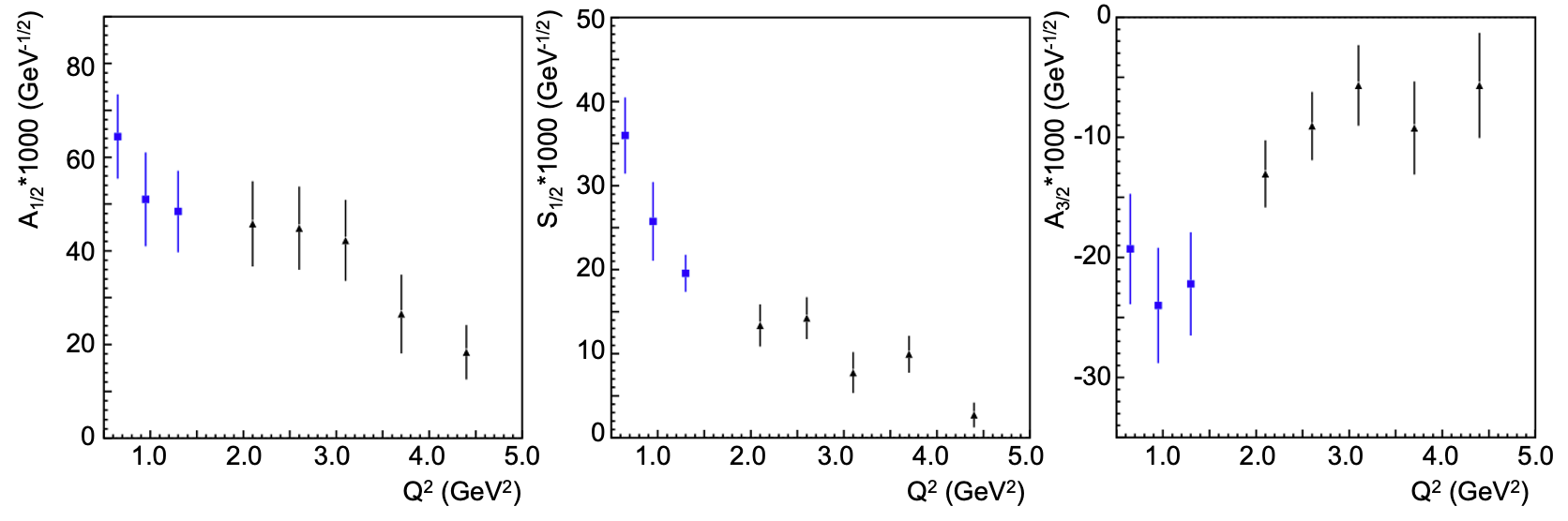}
\caption{Electrocouplings of the $N'(1720)3/2^+$, $A_{1/2}$ (left), $S_{1/2}$ (center), and $A_{3/2}$ (right), determined in the analyses of $\pi^+\pi^-p$ differential cross sections~\cite{Trivedi:2026fsd} (black) and the previous results for $Q^2 < 1.5$~GeV$^2$ \cite{Mokeev:2020hhu} (blue). The results for $Q^2 > 2.0$~GeV$^2$ are averaged across the two $W$ intervals $1.61-1.71$~GeV and $1.66-1.76$~GeV.}  
\label{Np1720elav}
\end{center}
\end{figure*}

The electrocouplings of the $N'(1720)3/2^+$ are shown in Fig.~\ref{Np1720el}. The electrocouplings averaged over the two $W$ intervals are shown in Fig.~\ref{Np1720elav} and their numerical values are listed in Table~\ref{Np1720el_final}. The non-resonant contributions differ across the two $W$ intervals. Nevertheless, the $N'(1720)3/2^+$ electrocouplings extracted from the data are consistent within their uncertainties across the entire $Q^2$ range. This consistency underscores the credible and robust extraction of these quantities.

The mass of the $N(1720)3/2^+$, along with its total decay width and partial decay widths to $\pi \Delta$ and $\rho p$, as determined from fits to the $\pi^+\pi^-p$ differential cross sections, are listed in Table~\ref{N1720hadr}. These values were obtained from four independent fits of the differential cross sections across overlapping $(W,Q^2)$ intervals: $1.61 < W < 1.71$~GeV and $1.66 < W < 1.76$~GeV, for $2.0 < Q^2 < 3.5$~GeV$^2$ and $3.0 < Q^2 < 5.0$~GeV$^2$.

\begin{figure*}[htbp]
\begin{center}
\includegraphics[width=0.9\textwidth]{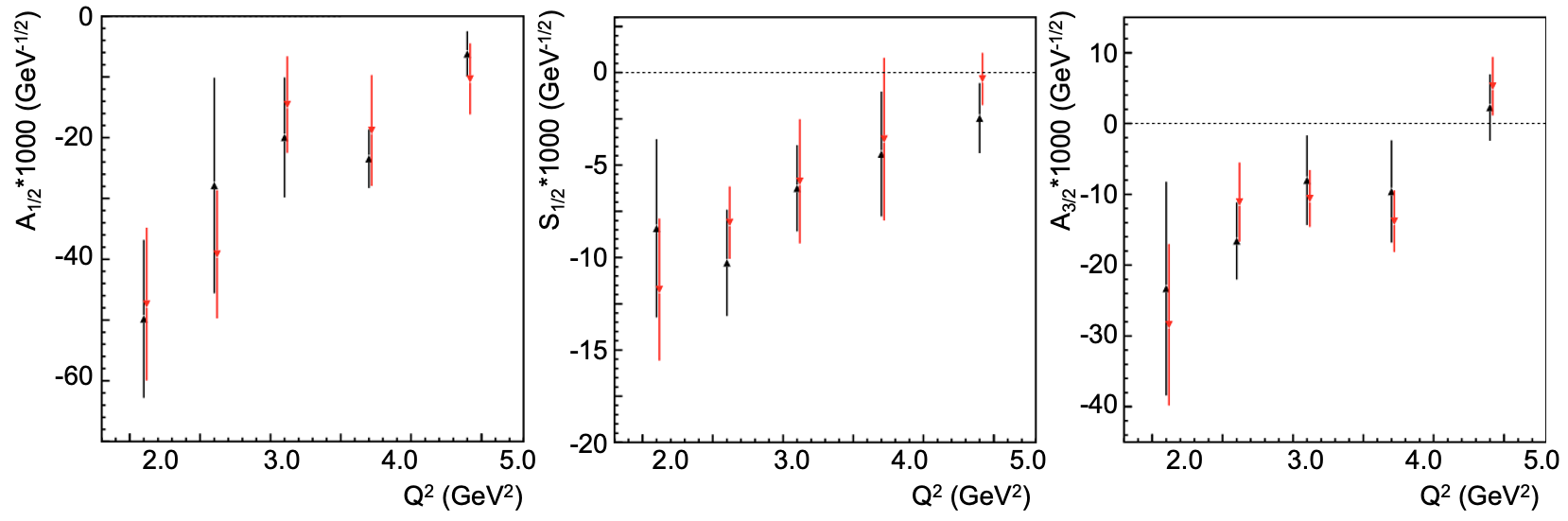}
\caption{Electrocouplings of the $N(1720)3/2^+$, $A_{1/2}$ (left), $S_{1/2}$ (center), and $A_{3/2}$ (right), determined in independent fits of $\pi^+\pi^-p$ CLAS data~\cite{Trivedi:2026fsd} across two overlapping $W$-intervals $1.61-1.71$~GeV (black) and $1.66-1.76$~GeV (red) for $Q^2$ from $2.0-5.0$~GeV$^2$.}  
\label{N1720el}
\end{center}
\end{figure*}

For comparison, Table~\ref{N1720hadr} also lists the $N(1720)3/2^+$ mass and hadronic decay parameters obtained in the previous analysis of $\pi^+\pi^-p$ photo- and electroproduction data~\cite{Mokeev:2020hhu}. The results of the present analysis available from independent fits of the $(W,Q^2)$ ranges listed in Table~\ref{N1720hadr} are consistent within the quoted parameter uncertainties. They are also consistent with the values previously extracted from the $\pi^+\pi^-p$ differential cross sections in the range $0.0 < Q^2 < 1.5$~GeV$^2$~\cite{Mokeev:2020hhu}. The $N(1720)3/2^+$ electrocouplings are shown in Fig.~\ref{N1720el}. Although the non-resonant amplitudes differ across the two overlapping $W$ intervals, the extracted $N(1720)3/2^+$ electrocouplings are coincident within uncertainties, providing strong evidence for the credibility of their extraction.

\begin{figure*}
\begin{center}
\includegraphics[width=0.9\textwidth]{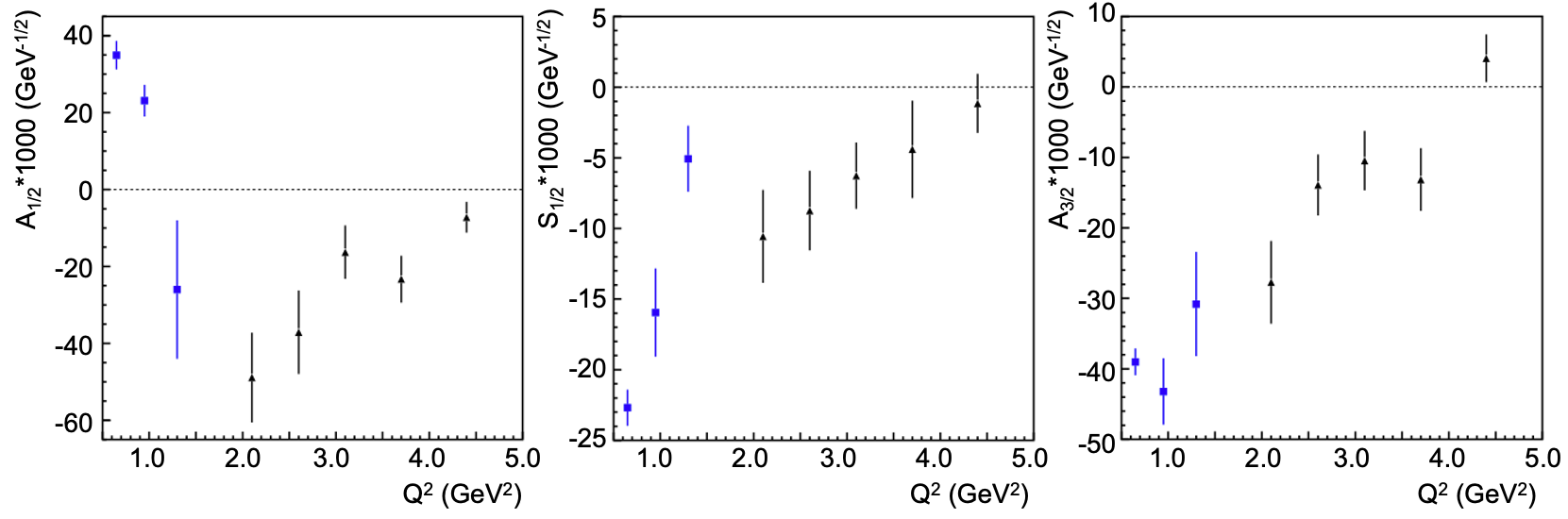}
\caption{Electrocouplings of the $N(1720)3/2^+$, $A_{1/2}$ (left), $S_{1/2}$ (center), and $A_{3/2}$ (right), determined in the analyses of $\pi^+\pi^-p$ differential cross sections~\cite{Trivedi:2026fsd} (black) and the previous results for $Q^2 < 1.5$~GeV$^2$ \cite{Mokeev:2020hhu} (blue). The results for $Q^2 > 2.0$~GeV$^2$ are averaged over the two $W$ intervals $1.61-1.71$~GeV and $1.66-1.76$~GeV.}  
\label{N1720elav}
\end{center}
\end{figure*}

The $N(1720)3/2^+$ electrocouplings, averaged over the two $W$ intervals, are shown in Fig.~\ref{N1720elav}, with the corresponding numerical values listed in Table~\ref{N1720el_final}. For comparison, the previously published results on the $N(1720)3/2^+$ electrocouplings are also shown in Fig.~\ref{N1720elav} (blue points). The $A_{1/2}$ electrocoupling exhibits a zero crossing at $Q^2 \approx 1.0$~GeV$^2$. As $Q^2$ increases, $A_{1/2}$ becomes negative, reaching a minimum near $Q^2 \approx 2.0$~GeV$^2$, before its absolute value decreases toward zero at $Q^2 \approx 4.5$~GeV$^2$. The presence of such a zero crossing at relatively large $Q^2$ is a distinctive feature in $N^*$ electroexcitation and may provide valuable input for theoretical approaches to the description of nucleon resonance structure.

The CSM has provided predictions for the wavefunction (Faddeev amplitude) of the $N(1720)3/2^+$~\cite{Liu:2022nku}. According to these studies, the dominant contribution ($\sim$94.1\%) to the $N(1720)3/2^+$ mass arises from diquark correlations with spin-parity $J^P = 1^+$. In contrast, the wavefunction composition is more complex, with the $J^P = 1^+$ diquark configuration contributing only 50\%. The remaining strength (44\%) primarily stems from the $J^P = 0^+$ diquark correlation, with subleading contributions from $J^P = 1^-$ configurations at the level of only a few percent.

In terms of orbital angular momentum contributions, the $N(1720)3/2^+$ mass in its rest frame is overwhelmingly dominated by quark-diquark relative orbital angular momentum $L_{qd} = 2$ ($D$-wave), which contributes 97.2\%. However, the detailed structure of the $N(1720)3/2^+$ wavefunction, as inferred from the canonically normalized Faddeev amplitude, is more intricate. It emerges from the interplay between $L_{qd} = 2$ and $L_{qd} = 3$ configurations, underscoring the complexity of the internal dynamics of this state.

These findings demonstrate that studies based solely on the resonance spectrum are insufficient to reveal the full complexity of the internal structure of the excited states of the nucleon. In particular, the $N(1720)3/2^+$ mass alone is not sensitive to the intricate structure of its wavefunction. Therefore, the results presented in this paper on the $Q^2$ evolution of the $N(1720)3/2^+$ electrocouplings over a broad range up to 5.0~GeV$^2$ offer a promising opportunity to gain insight into the detailed structure of the $N(1720)3/2^+$ wavefunction predicted by the CSM and other theoretical approaches. Furthermore, within the CSM framework, the $N(1520)3/2^-$ and $N(1720)3/2^+$ are regarded as chiral partners. Hence, comparative studies of the CLAS results on the $Q^2$ evolution of their electrocouplings can illuminate the deep connection between EHM and DCSB.

Analysis of the CLAS data on the $\pi^+\pi^-p$ differential cross sections in the third resonance region~\cite{Trivedi:2026fsd}, performed within the JM23 meson-baryon reaction model~\cite{Mokeev:2023zhq}, conclusively demonstrated contributions from both the conventional $N(1720)3/2^+$~\cite{ParticleDataGroup:2024cfk} and the new $N'(1720)3/2^+$~\cite{Mokeev:2020hhu} resonances. These two excited states have close masses and identical spin-parity quantum numbers, $J^P = 3/2^+$, but exhibit distinct decay patterns into $\pi\Delta$ and $\rho p$, which prevents their mixing (see Tables~\ref{Np1720hadr} and~\ref{N1720hadr}). The evolution of the electrocouplings for the $N'(1720)3/2^+$ and $N(1720)3/2^+$ with $Q^2$, shown in Fig.~\ref{Np1720elav} and Fig.~\ref{N1720elav}, is markedly different, enabling the disentanglement of their individual contributions in the fit to the $\pi^+\pi^-p$ differential cross sections.

\begin{table*}[htbp]
\begin{center}
\begin{tabular}{|c|c|c|c|c|c|c|} \hline
                                                & $\Gamma_\text{tot}$ (MeV) & $\Gamma_{\pi\Delta}$ (MeV) & BF($\pi\Delta$) (\%) & $\Gamma_{\rho p}$ (MeV) & BF($\rho p$) (\%) & Mass (GeV) \\   \hline
$W$: $1.61-1.71$ GeV, $Q^2$: $2.0-3.5$ GeV$^2$  & $112.6 \pm 2.9$     & $49.1 \pm 3.4$            & $39-48$               & $44.0 \pm 3.3$        & $35-43$          & $1.739 \pm 0.004$ \\ \hline     
$W$: $1.61-1.71$ GeV, $Q^2$: $3.0-5.0$ GeV$^2$  & $114.3 \pm 4.2$     & $51.2 \pm 1.5$            & $42-48$               & $43.7 \pm 5.3$        & $33-45$          & $1.738 \pm 0.007$ \\ \hline
$W$: $1.66-1.76$ GeV, $Q^2$: $2.0-3.5$ GeV$^2$  & $118.0 \pm 3.5$     & $51.0 \pm 3.3$            & $39-47$               & $47.6 \pm 4.5$        & $35-45$          & $1.736 \pm 0.006$ \\ \hline
$W$: $1.66-1.76$ GeV, $Q^2$: $3.0-5.0$ GeV$^2$  & $117.7 \pm 4.5$     & $51.5 \pm 3.5$            & $39-49$               & $46.8 \pm 3.9$        & $35-45$          & $1.741 \pm 0.012$ \\ \hline
$W$: $1.61-1.76$ GeV, $Q^2$: $0.0-1.5$ GeV$^2$  & $114 \pm 6$         &                           & $39-53$               &                       & $31-46$          & $1.748 \pm 0.005$ \\ \hline
\end{tabular}
\caption{The mass, total decay width ($\Gamma_\text{tot}$), partial decay widths ($\Gamma_{\pi\Delta}$, $\Gamma_{\rho p}$), and branching fractions (BF$_{\pi\Delta}$, BF$_{\rho p}$) for the $N(1720)3/2^+$ determined from independent fits of the nine one-fold $\pi^+\pi^-p$ differential cross sections \cite{Trivedi:2026fsd}. The results from our previous publication~\cite{Mokeev:2020hhu} with the $N(1720)3/2^+$ parameters determined from $\pi^+\pi^-p$ photo- and electroproduction differential cross sections for $Q^2$ from the photon point up to 1.5~GeV$^2$ are shown for comparison in the bottom row.}
\label{N1720hadr} 
\end{center}
\end{table*}

A good description of the $\pi^+\pi^-p$ differential cross sections across the third resonance region has been achieved using $Q^2$-independent masses and total as well as partial hadronic decay widths to $\pi\Delta$ and $\rho p$ for both the $N'(1720)3/2^+$ and $N(1720)3/2^+$ (see Tables~\ref{Np1720hadr} and~\ref{N1720hadr}). This result suggests that both states, excited in the $s$-channel of the virtual photon-proton interaction, possess cores of three dressed quarks surrounded by meson-baryon clouds. Furthermore, the successful description of the data over a broad range of $Q^2$, from the real photon point up to $Q^2 = 5.0$~GeV$^2$, using $Q^2$-independent masses and total/partial decay widths not only for the $N'(1720)3/2^+$ but also for all other resonances contributing to the $\pi^+\pi^-p$ cross sections in the third resonance region, provides strong and nearly model-independent evidence for the existence of the new $N'(1720)3/2^+$. If, instead, the inclusion of the $N'(1720)3/2^+$ was merely an effective way to compensate for shortcomings of the JM23 model in describing the non-resonant amplitudes, the $\pi^+\pi^-p$ cross sections could not be reproduced with $Q^2$-independent mass and decay parameters for this resonance, due to the pronounced dependence of the non-resonant amplitudes on $Q^2$.

The new results presented in this paper provide solid evidence for the existence of the $N'(1720)3/2^+$. Earlier, the global multi-channel analysis of photo- and hadroproduction data by the Argonne-Osaka group~\cite{Kamano:2013iva} revealed two closely spaced $N^*$ states with spin-parity $J^P = 3/2^+$ located within the mass range $1.7-1.8$~GeV. The presence of two $N^*$ states with $J^P = 3/2^+$ in this mass interval had also been predicted within the relativistic interacting quark-diquark model~\cite{Santopinto:2014opa} and the hypercentral constituent quark model~\cite{Giannini:2015zia}. The studies of $\pi^+\pi^-p$ differential cross sections in the third resonance region presented here, as well as in previous work~\cite{Mokeev:2020hhu}, conclusively confirm the findings of the Argonne-Osaka coupled-channel analysis and are consistent with the quark model predictions. 

At present, the $N'(1720)3/2^+$ is the only newly established baryon state for which information on the $Q^2$-evolution of its electrocouplings has become available over a broad $Q^2$ range up to 5.0~GeV$^2$, based on measurements of the $\pi^+\pi^-p$ differential cross sections with CLAS~\cite{CLAS:2018drk, CLAS:2002xbv, CLAS:2017fja, Trivedi:2026fsd}. An anti-de Sitter (AdS) /QCD-based model has been developed~\cite{Lyubovitskij:2020gjz}, enabling the determination of contributions from the relevant AdS fields to the $N^*$ structure by fitting to the experimental results on the $Q^2$-evolution of the electrocouplings. The detailed information on the $N'(1720)3/2^+$ electrocouplings for $Q^2 < 5.0$~GeV$^2$ offers a promising avenue to explore the internal structure of this newly discovered resonance, shedding light on the distinctive features of the so-called ``missing'' resonances that have long evaded experimental observation.

The first results from $\pi^+\pi^-p$ electroproduction on the $Q^2$-evolution of the $\gamma_v p N^*$ electrocouplings for the $N(1675)5/2^-$, $N(1680)5/2^+$, $\Delta(1700)3/2^-$, $N(1720)3/2^+$, and $N'(1720)3/2^+$ resonances in the range $2.0 < Q^2 < 5.0$~GeV$^2$ provide valuable input for the development of constituent quark models that describe the internal structure of $N^*$ states in the mass region $1.6-1.75$~GeV. The need to extend the experimental information on the electrocouplings to $Q^2 > 2.0$~GeV$^2$ has been emphasized in a recent review~\cite{Ramalho:2023hqd}.

\begin{table}[htbp]
\begin{center}
\begin{tabular}{|c|c|c|c|}
\hline
$Q^2$ Interval & $A_{1/2} \times 1000$ & $S_{1/2} \times 1000$ & $A_{3/2} \times 1000$  \\
 (GeV$^2$)     & (GeV$^{-1/2}$)        & (GeV$^{-1/2}$)        & (GeV$^{-1/2}$)  \\ \hline
 $2.0-2.4$     & $-48.4 \pm 11.7$      & $-10.6 \pm 3.3$       & $-27.7 \pm 11.1$ \\ \hline
 $2.4-3.0$     & $-37.1 \pm 10.8$      & $-8.7 \pm 1.8$        & $-13.9 \pm 4.3$ \\ \hline
 $3.0-3.5$     & $-16.3 \pm 7.0$       & $-6.3 \pm 2.4$        & $-10.5 \pm 4.2$ \\ \hline
 $3.5-4.2$     & $-23.3 \pm 6.1$       & $-4.4 \pm 3.5$        & $-13.1 \pm 4.5$ \\ \hline
 $4.2-5.0$     & $-7.2 \pm 4.0$        & $-1.2 \pm 1.4$        & $4.1 \pm 3.4$ \\ \hline
\end{tabular}
\caption{$N(1720)3/2^+$ electrocouplings determined from fits of the $\pi^+\pi^-p$ differential cross sections \cite{Trivedi:2026fsd} and averaged over the $W$ intervals $1.61-1.71$~GeV and $1.66-1.76$~GeV for $Q^2$ from $2.0-5.0$~GeV$^2$.} 
\label{N1720el_final}
\end{center}
\end{table}

\section{Conclusions and Outlook}
\label{concl_outlook}

A successful description of the nine one-fold $\pi^+\pi^-p$ differential cross sections across the $W$ range of $1.56-1.76$ GeV for $Q^2$ from $2.0-5.0$~GeV$^2$ has been achieved within the JM23 reaction model~\cite{Mokeev:2023zhq}. This achievement enabled a credible separation between the resonant and non-resonant contributions, allowing for the extraction of the $\gamma_v p N^*$ electrocouplings of the $N(1675)5/2^-$, $N(1680)5/2^+$, $\Delta(1700)3/2^-$, $N'(1720)3/2^+$, and $N(1720)3/2^+$ for the first time in this kinematic domain. The reliability of the extracted electrocouplings has been confirmed by the consistent results obtained for their $Q^2$ evolution from independent fits in overlapping $W$ intervals.

Furthermore, the electrocouplings of the $N(1675)5/2^-$ and $N(1680)5/2^+$ determined from independent studies of $\pi N$~\cite{Park:2014yea} and $\pi^+\pi^-p$ electroproduction reported in this work are consistent within their quoted uncertainties. Since the $\pi N$ and $\pi^+\pi^-p$ electroproduction channels are the dominant contributors in the resonance region, yet involve different non-resonant amplitudes, the observed consistency of the extracted electrocouplings across the broad $Q^2$ range provides strong evidence for the reliability of the reaction models developed by the CLAS Collaboration in enabling their determination.

From the fits to the $\pi^+\pi^-p$ differential cross sections, the total and partial decay widths to $\pi\Delta$ and $\rho p$ for the $N(1675)5/2^-$, $N(1680)5/2^+$, $\Delta(1700)3/2^-$, $N'(1720)3/2^+$, and $N(1720)3/2^+$ are found to be $Q^2$-independent from the photon point to $Q^2 = 5.0$~GeV$^2$. This behavior suggests that each of these states are excited in the $s$-channel of the virtual photon-proton interaction and possess inner cores composed of three dressed quarks, which may be surrounded by an external meson-baryon cloud.

The analysis of the $\pi^+\pi^-p$ electroproduction data within the JM23 model~\cite{Mokeev:2023zhq} for $1.61 < W < 1.76$~GeV and $2.0 < Q^2 < 5.0$~GeV$^2$ has revealed the presence of two nearby resonances, $N(1720)3/2^+$ and $N'(1720)3/2^+$. These states have identical spin-parities and closely spaced masses; however, they exhibit distinctly different hadronic decay patterns into $\pi\Delta$ and $\rho p$ (see Tables~\ref{Np1720hadr} and~\ref{N1720hadr}), which prevents their mixing. Their $Q^2$ evolution also differs significantly (see Figs.~\ref{Np1720elav} and~\ref{N1720elav}), enabling a clear separation of their respective contributions to the differential cross sections.

A satisfactory description of the $\pi^+\pi^-p$ differential cross sections obtained in the present analysis, as well as in our previous study~\cite{Mokeev:2020hhu}, has been achieved using $Q^2$-independent masses, and total and partial hadronic decay widths to $\pi\Delta$ and $\rho p$ for both the $N(1720)3/2^+$ and $N'(1720)3/2^+$ over the range from the photon point to $Q^2 = 5.0$~GeV$^2$. This result provides strong evidence for the existence of the new $N'(1720)3/2^+$ state. To date, the $N'(1720)3/2^+$ is the only newly established resonance for which information on the $Q^2$ evolution of its electrocouplings has become available. These results offer valuable insight into the distinctive features of the so-called ``missing'' resonance structures that have long hindered their experimental identification.

The new experimental results on the electrocouplings of chiral-partner excited states of the nucleon-$\Delta(1232)3/2^+$ and $\Delta(1700)3/2^-$, $N(1520)3/2^-$ and $N(1720)3/2^+$, $N(1675)5/2^-$ and $N(1680)5/2^+$-open new opportunities to investigate the manifestations of DCSB in connection with EHM within the CSM framework~\cite{Achenbach:2025kfx, Ding:2022ows, Albino:2025fcp}. The successful description of the $\Delta(1232)3/2^+$ and $\Delta(1700)3/2^-$ chiral-partner state electrocouplings within CSM \cite{Segovia:2014aza, Cheng:2025sdp} demonstrated the connection between DCSB and EHM for the first time in baryon sector. The results on the electrocouplings for resonances in the mass range $1.60-1.76$~GeV and $Q^2$ from $2.0-5.0$~GeV$^2$ enhance our ability to test the universality-or reveal the environmental sensitivity-of the dressed-quark mass function, as well as to explore the nature of diquark correlations with different spin-parity configurations. Ultimately, these studies will provide critical input for establishing the partial wave decomposition of the $qq$ scattering amplitudes describing diquark correlations inside the ground and excited states of the nucleon.

Resonance physics remains a challenge for the numerical simulation of lattice regularized QCD, but progress is being made~\cite{Chen:1980qh, Morningstar:2025khf}. Furthermore, the new results on the electrocouplings presented in this paper are expected to stimulate further development of quark models~\cite{Ramalho:2023hqd, Burkert:2025coj}. These advances will open new avenues for describing the structure of $N^*$ states composed of three dressed quarks in various orbital and radial excitations, as well as for exploring the interplay between the inner core of three dressed quarks and the external meson-baryon cloud based on the empirical information from the electrocoupling data.

\begin{acknowledgments}
The authors would like to acknowledge the outstanding efforts of the Physics Division staff at JLab that made this analysis possible. The authors express their gratitude to Prof. C.D. Roberts for reviewing this manuscript and for providing feedback. This work was supported in part by the U.S. Department of Energy (DOE), Office of Science, Office of Nuclear Physics under contract 89243126CSC000213, the National Science Foundation (NSF) under Grant PHY 1812382, the Physics Department of the University of South Carolina (USC) under NSF Grant PHY 10011349, and the Skobeltsyn Nuclear Physics Institute and Physics Department at Lomonosov Moscow State University.
\end{acknowledgments}

\nocite{*}
\bibliography{References}

\end{document}